\documentclass[amssymb,amsmath,twocolumn,superscriptaddress,nofootinbib,floatfix]{revtex4-2}
\usepackage[table, dvipsnames, xcdraw]{xcolor}
\usepackage{nicematrix}
\usepackage{graphicx}
\usepackage{bm}
\usepackage{amssymb,amsmath}
\usepackage{mathrsfs}
\usepackage{latexsym}
\usepackage[normalem]{ulem}
\usepackage{dcolumn}
\usepackage{orcidlink}

\usepackage{graphicx}
\usepackage{soul}
\usepackage[commentmarkup=uwave]{changes}
\usepackage{multirow, tabularray}
\usepackage{booktabs}
\graphicspath{{./}{figures/}}

\allowdisplaybreaks

\newcommand{\CIT}{\affiliation{TAPIR, California Institute of Technology, Pasadena, CA 91125, USA}}
\newcommand{\CITLab}{\affiliation{LIGO Laboratory, California Institute of Technology, Pasadena, California 91125, USA}}

\makeatletter
\newcommand*{\rom}[1]{\expandafter\@slowromancap\romannumeral #1@}
\makeatother

\definecolor{kcmagenta}{rgb}{0.54, 0.17, 0.88}
\definecolor{shyellow}{rgb}{0.15625, 0.609375, 0.316406}
\definecolor{TableGrayLight}{gray}{0.98} 
\definecolor{TableGrayDark}{gray}{0.85}  

\newcommand{\BW}{\texttt{BayesWave}}
\newcommand{\BWCpp}{\texttt{BayesWaveCpp}}

\newcommand{\chieff}{\chi_{\textrm{eff}}}
\newcommand{\chip}{\chi_\mathrm{p}}

\newcommand{\Hz}{\mathrm{Hz}}

\begin{document}

\title{Glitches far from transient gravitational-wave events do not bias inference}

\author{Sophie Hourihane
\orcidlink{0000-0002-9152-0719}}
\email{sohour@caltech.edu}\CIT \CITLab

\author{Katerina Chatziioannou 
\orcidlink{0000-0002-5833-413X}}
\email{kchatziioannou@caltech.edu}\CIT \CITLab 
\date{\today}

\begin{abstract}
Non-Gaussian noise in gravitational-wave detectors, known as ``glitches,'' can bias the inferred parameters of transient signals when they occur nearby in time and frequency. 
These biases are addressed with a variety of methods that remove or otherwise mitigate the impact of the glitch.
Given the computational cost and human effort required for glitch mitigation, we study the conditions under which it is strictly necessary. 
We consider simulated glitches and gravitational-wave signals in various configurations that probe their proximity both in time and in frequency. 
We determine that glitches located outside the time-frequency space spanned by the gravitational-wave model prior and with a signal-to-noise ratio, conservatively, below 50 do not impact estimation of the signal parameters.
\end{abstract}

\maketitle

\section{Introduction}\label{sec:intro}
 
The properties of compact binary coalescences (CBCs) observed via gravitational-waves (GWs)~\cite{gwtc3, O3a_catalogue}, offer insights on the astrophysical properties of black holes and neutron stars~\cite{KAGRA:2021duu} and test General Relativity~\cite{LIGOScientific:2021sio}.
Since the first direct detection of GWs~\cite{GW150914} the LIGO-Virgo-Kagra (LVK) detector network~\cite{LIGOScientific:2014pky, VIRGO:2014yos} has identified 89 more CBCs~\cite{gwtc3}.
The current fourth observing run has so far tallied a further ${\sim}200$ event candidates~\cite{LIGO_Public_Alerts}, for
an approximate event rate of once every other day. 
 
Also present in the data are transient, short-duration bursts of power of terrestrial origin, ``glitches,'' with a rate of 0.5-1.28 per minute \textit{in each detector}~\cite{LIGO:2024kkz_detcharO4}. 
When glitches overlap with a signal, i.e., they are coincident in both time and frequency, they impact the inferred source parameters~\cite{Hourihane:2022doe_glitch, Ghonge:2023ksb, Pankow:2018qpo, Payne:2022spz_200129, Udall:2024ovp_191109, Macas:2022afm, Powell:2018csz, Kwok:2021zny, Mozzon:2021wam, Chatziioannou:2021ezd}.
This is because glitches violate two fundamental assumptions that underlie GW inference, namely that the noise is stationary and Gaussian. 
In practice, the impact of glitches ranges from biasing inference of subdominant effects like spin-precession~\cite{Payne:2022spz} or spin-alignment~\cite{Udall:2024ovp_191109}, to mimicking an entirely different signal~\cite{TheLIGOScientific:2017qsa,Pankow:2018qpo}.
Simply put, glitches make the noise non-Gaussian such that the common Whittle likelihood no longer describes it. 

GW inference is based on a time-frequency ``analysis window'' that is determined by the detector and signal properties, see Appendix E of~\citet{gwtc3} for details. 
The window frequency extent is defined by a lower bound,  $f_{\mathrm{low}} =20\,\mathrm{Hz}$, set by the detector low-frequency sensitivity, and an upper bound, $f_{\mathrm{high}}$, chosen to contain the merger frequency of the $(\ell, |m|) = (3,3)$ signal mode. 
The time extent is designed to enclose the signal from $f_{\mathrm{low}}$ to merger (and ringdown) with an additional 2\,s of data post-merger~\cite{GWTC2.1}. 
For reference, a typical analysis window for a 30\,$M_\odot$+30\,$M_\odot$ (detector-frame masses) binary is $4-8\,$s whereas for a 1.4\,$M_\odot$+1.4\,$M_\odot$ binary the analysis window is $128-256\,$s in length.

Every event candidate is vetted for glitches within the analysis window~\cite{LIGO:2024kkz_detcharO4, Davis:2022ird}, and if one is identified, it is subtracted~\cite{Davis:2022ird}. 
Notably, this procedure flags glitches anywhere in the analysis window, regardless of whether they overlap, with the signal, i.e., whether they intercept the actual signal time-frequency track~\cite{LIGO:2024kkz_detcharO4, Davis:2022ird}. 
For instance, of the 16 events flagged for glitch subtraction in the third LVK observing period~\cite{gwtc3, O3a_catalogue},\footnote{An additional 10 were initially flagged by visual inspection, but were ultimately deemed consistent with Gaussian noise~\cite{Davis:2022ird,Vazsonyi:2022jul}.} 
in only 12 cases did the glitch overlap with the dominant, $(\ell, |m|) = (2,2)$, signal mode.
Since each event requires considerable compute and person time (including vetting, review, run time, etc.~\cite{Davis:2022ird}) it is desirable to restrict to only cases were mitigation is necessary to avoid biases.  

\begin{figure}
    \centering
    \includegraphics[width=\linewidth]{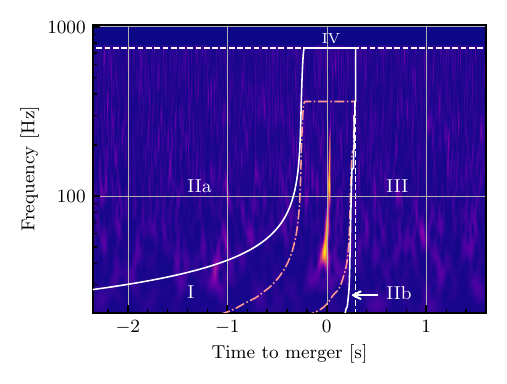}
    \caption{
    Time-frequency breakdown of the CBC analysis window, schematically describing the ways in which glitches can be positioned with respect to signals. 
    Region \rom{1} (solid lines) encloses all time-frequency tracks (including those of higher order modes) within the analysis prior. For reference, in orange we plot the time-frequency content of the (2, 2) mode prior.
    Region \rom{2} contains glitches coincident in time and frequency, but never concurrently. We split between Region \rom{2}a, those above the (4,4) mode in frequency, and Region \rom{2}b, those below the (2,1) mode. 
    Region \rom{3} contains glitches coincident in frequency, but not in time. 
    Region \rom{4} contains glitches not coincident in frequency.  
   }
    \label{fig:schematic}
\end{figure}

In this study, we explore the conditions under which glitch mitigation can be avoided.  
Specifically, we address \textit{whether glitches that do not overlap with signals in time-frequency bias inference}. 
We focus on high-detector-frame mass events, $m \in (20, 100)\,M_\odot$, as glitches disproportionately impact shorter, higher-mass events compared to longer ones containing neutron stars~\cite{Hourihane:2022doe_glitch}. 
This is likely due to the fact that inference for short signals hinges on less data, sometimes a single cycle~\cite{Payne:2022spz, Miller:2023ncs}, and can thus be affected by short glitches. 
We schematically lay out the potential glitch-signal configurations in Fig.~\ref{fig:schematic} that outlines four (non exclusive) regions. 
 
\begin{itemize}
    \item \textbf{Region \rom{1}} contains glitches that overlap with \textit{at least one signal within the CBC model prior}. 
    Previous studies~\cite{Hourihane:2022doe_glitch, Ghonge:2023ksb, Pankow:2018qpo, Payne:2022spz_200129, Udall:2024ovp_191109, Macas:2022afm, Powell:2018csz, Kwok:2021zny, Mozzon:2021wam, Chatziioannou:2021ezd} have shown that such overlaps can lead to biases when not mitigated properly.
    We confirm these results, primarily using this region for comparison to others. 
    \item \textbf{Region \rom{2}} contains glitches that are coincident in time \textit{and} frequency, but not simultaneously. 
    We further distinguish between Region \rom{2}a, glitches above the signal, and Region \rom{2}b, glitches below the signal (in frequency). 
    Since we are focusing on high-mass events that evolve faster and merge at comparatively lower frequencies than their lower-mass counterparts, Region \rom{2}b is very small. 
    We therefore focus on the comparatively much larger Region \rom{2}a. 
    We evaluate  differences in CBC parameter posteriors  from data with and without glitches within this region. Glitches can induce a bias only when they have a signal-to-noise ratio (SNR) (conservatively) above 50 \textit{and}  they are close in frequency to the CBC merger frequency.
    \item \textbf{Region \rom{3}} includes glitches that share frequency content with the signal, but are not coincident in time. 
    We again simulate glitches within that region and vary the distance (in time) from the signal, as well as the glitch SNR.\footnote{We only consider glitches \textit{after} the signal, as those before it are typically cut from the analysis window.}
    Via a standard P-P test, we find that glitches in this region leave no statistical imprint on a population of signals.
    When considering individual glitches and a GW150914-like signal~\cite{GW150914}, glitches with SNR $<50$ never induce a bias, neither do glitches with SNR$<100$ if they occur more than 0.5\,s after the signal. 
    \item \textbf{Region \rom{4}} contains glitches that do not share any frequency content with the event.
    We analytically show that such glitches do not impact inference.
\end{itemize}

The rest of the paper is organized as follows. 
In Sec.~\ref{sec:background} we recap current glitch mitigation techniques and lay out the noise assumptions that are the foundations for CBC parameter estimation. 
In Sec.~\ref{sec:methods} we lay out the methodology of our study. 
In Sec.~\ref{sec:results} we go through each Region in Fig.~\ref{fig:schematic} and present our results. 
In Sec.~\ref{sec:conclusion} we conclude.

\section{Background}~\label{sec:background}

In this section we recap glitch mitigation in Sec.~\ref{sec:GlitchMitigation} and standard inference under Gaussian noise in Sec.~\ref{sec:whittle}.

\subsection{Glitches and Glitch Mitigation}\label{sec:GlitchMitigation}
 
During the third observing run, glitches occurred more than once every minute in each detector~\cite{LIGOScientific:2020ibl, gwtc3}.
Of the total of 79 events with an astrophysical probability greater than 0.5, 16 contained glitches within their analysis window.
In 4 of 16 glitch-mitigated events, the glitch did \textit{not} overlap the signal time-frequency track~\cite{Davis:2022ird, gwtc3, GWTC2.1, O3a_catalogue}.
Assuming a fixed glitch rate between $1-1.5$ glitches per minute, the number of signals requiring glitch mitigation during the fourth LVK observing run is expected to increase simply due to the increased event rate.
Adopting a 4\,s analysis window and two detectors, we expect 26-40 ($13-20\%$) of the 203 current candidates to contain glitches within their analysis window~\cite{LIGO_Public_Alerts}.
The increased demand for glitch mitigation motivates our detailed look into the conditions that require it.

The amount of glitch-signal overlap will determine the degree to which inference is biased.
A glitch and a GW signal overlap in frequency when, under a Fourier decomposition, there are frequencies, $f_i$, for which they are both nonzero. 
A glitch and a signal overlap in time when there are times, $t_i$, for which they are both nonzero. 
A glitch and a signal overlap in time-frequency if, when decomposed into time-frequency space, there are bins with non-zero content from both. 
Therefore a glitch and signal can overlap in both time and frequency, but \textit{not} overlap in time-frequency.  
These definitions apply even for GWs with higher-order modes, for which there is no 1-1 relationship between time and frequency. 
In this case, we consider all CBC modes $(\ell, |m|)$ to determine overlaps, e.g., Fig.~\ref{fig:schematic}.

Currently, if a glitch cannot be excluded from the analysis by redefinition of the analysis window (subject to the established criteria~\cite{gwtc3}), it is subject to mitigation. 
Common mitigation methods include: (i) subtracting a single estimate of the glitch from the data, a process known as ``glitch subtraction''~\cite{Davis:2018yrz, Davis:2022ird}, (ii) modeling both the signal and the glitch and thus marginalizing over the uncertainty of both models~\cite{Hourihane:2022doe_glitch, Ghonge:2023ksb, Chatziioannou:2019zvs, Udall:2024ovp_191109}, and (ii) removing all affected data by zeroing~\cite{Zweizig:gating} or by replacing with Gaussian noise~\cite{Zackay:2019kkv}. 
Glitch subtraction requires some (time or frequency) reconstruction of the glitch.
Some glitch classes can be described with physical models such as slow and fast scattering~\cite{Soni:2023kqq, LIGO:2020zwl,
Udall:2022vkv}.  Another option is to utilize auxiliary data channels that ``witness'' noise. By determining the transfer function between these witness channel(s) and the measured strain data, it is possible to subtract the associated glitch~\cite{Davis:2018yrz}, but only if such channels exist. 
In the absence of  physically-motivated models or witness channels, most glitches are targeted phenomenologically with $\BW$~\cite{Cornish:2020dwh, Cornish:2014kda}, further described in Sec.~\ref{sec:bayeswave}.

\subsection{Gaussian Noise likelihood}\label{sec:whittle}

The noise assumptions underlying GW data analysis determine the form of the likelihood~\cite{Romano:2016dpx}. 
The data, $\mathbf{d}$, are a combination of a GW signal $\mathbf{h}$ and noise $\mathbf{n}$, which is a sum of Gaussian, $\mathbf{n_{G}}$, and transient, non-Gaussian noise (glitches), $\mathbf{g}$. 
All quantities are considered in the frequency domain. 
For stationary $\mathbf{n_G}$, the noise is uncorrelated between frequency bins, meaning that the frequency domain noise-covariance matrix is diagonal. 
Gaussianity means that the noise is described by a Gaussian distribution at each frequency. 
The per-frequency Gaussian distribution is then entirely described by the variance, leading to a frequency-domain noise-covariance matrix that is proportional to the noise power spectral density (PSD), $\mathbf{S_n}$. 

The likelihood for the GW model with parameters $\theta$, $\mathbf{h}_\theta$ in a single detector is then~\cite{Romano:2016dpx},
\begin{align}\label{eq:likelihood}
     \mathcal{L}\left(\mathbf{d} | \theta \right)&= \exp \left( -\frac{1}{2}|\mathbf{d} -\mathbf{h}_\theta|^2 -\sum_{f_i}\ln \left(2\pi S_{ni}\right)\right)\,,
\end{align}
where $|\mathbf{a}|$ is the noise-weighted magnitude
\begin{equation}
    |\mathbf{a}| = \sqrt{(\mathbf{a} | \mathbf{a})}\,,
\end{equation} 
$(\mathbf{a} | \mathbf{b})$ is the noise-weighted inner product
\begin{equation}
    (\mathbf{a} | \mathbf{b}) = 4 \Delta f\sum_{f_i = f_\mathrm{low}}^{f_\mathrm{high}} \frac{a_i {b_i}^*}{S_{ni}}\,,
\end{equation}
and $\Delta f$ is the frequency resolution.
The likelihood across all detectors is the product of each individual-detector likelihood. 

\section{Methods}~\label{sec:methods}

In this section, we describe our methodology.
We simulate data containing CBC signals and glitches and analyze them ignoring the presence of the glitch.
In Sec.~\ref{sec:bayeswave} we introduce the $\BWCpp$ analysis package that is used to  sample from the posterior for the CBC parameters, Sec.~\ref{sec:CBC_model}, and to simulate glitches, Sec~\ref{sec:glitch_model}. 
We then introduce ``glitch reweighting'' in Sec.~\ref{sec:reweighting} as a method to (i) quickly generate posteriors and (ii) to quantify the degree of similarity between two probability distributions, specifically those obtained from data with and without glitches.
We introduce the Jensen Shannon Divergence in Sec.~\ref{sec:JSD} as another quantity with which to compare two posteriors.

\subsection{BayesWaveCpp}\label{sec:bayeswave}

$\BWCpp$~\cite{bayeswave_cpp} is a rewrite of and upgrade to $\BW$~\cite{Cornish:2014kda,Cornish:2020dwh, bayeswave}, a software package used to stochastically sample the posteriors of signals, glitches, and noise PSDs in GW data. 
GWs can be modeled through coherent sums of sine-Gaussian wavelets, physical CBC waveform models, and combinations thereof. 
Glitches are modeled as sums of sine-Gaussian wavelets.
The noise PSD is modeled via broadband splines and Lorentzians for the spectral lines. 
For the purposes of this study, $\BWCpp$ is only used to sample the CBC posterior under a CBC waveform model. 
Though it has the capacity to do so, we do not sample the glitch or PSD posterior, assuming no glitch for the former and a known PSD for the latter. 

\subsubsection{CBC model}\label{sec:CBC_model}
\begin{table}[]
    \centering
    \begin{tblr}{
        colspec  = {lcl},
        rowsep   = 3pt,
        row{odd} = {bg=TableGrayDark},
        row{even} = {bg=TableGrayLight},
    }  
        \textbf{Parameter} & \textbf{Symbol} & \textbf{Prior} \\
        \hline
        Mass & $m_i$ & U[20, 100]\,$M_\odot$  \\
        Spin amplitude & $\chi_i$ & U[0, 1] \\ 
        Spin in-plane angle & $\phi_i$ & U[0, 2$\pi$] \\ 
        Spin polar angle & $\theta_i$ & $\cos{\theta_i} \sim$ U[-1, 1] \\ 
        Hanford time & $t_{\mathrm{LHO}}$ & U[-0.05, 0.05]\,s \\ 
        Luminosity distance & $D_L$ & $D_L^3 \sim$ U[1, $10000^3$]\,$\mathrm{Mpc}^3$ \\ 
        Inclination & $\iota$ & $\cos{\iota} \sim$ U[0, 1] \\ 
        Right ascension & $\alpha$ & U[0, 2$\pi$] \\ 
        Declination & $\delta$ & $\sin{\delta} \sim$ U[-1, 1] \\ 
        Polarization & $\psi$ & U[0, 2$\pi$] \\ 
        Coalescence phase & $\phi$ & U[0, 2$\pi$] \\ 
    \end{tblr}
    \caption{Parameter definition, notation, and prior distribution for all CBC parameters. Here $i \in \{1,2\}$ indexes the compact objects; $i=1$ (2) is the larger (smaller) mass object. The time in the Hanford detector is centered $2\,$s before the end of the $4\,$s analysis window.}
    \label{tab:CBCPrior}
\end{table}

In this study, we model all CBC signals with $\texttt{IMRPhenomXPHM}$~\cite{Pratten:2020ceb}, an inspiral-merger-ringdown model that includes precession and higher-order modes, but does not include eccentricity.\footnote{The ability to model precession is novel to $\BWCpp$, whereas $\BW$ is restricted to spin-aligned waveforms.} 
The CBC parameters and their priors are listed in Table~\ref{tab:CBCPrior}, here we briefly describe some parameters of interest that feature in the following figures. 
The total mass, $M$, is the sum of the component masses of the compact binary in the detector frame. 
The spin angular momentum of the compact binary system can be summarized with ``effective" parameters: the effective precessing parameter $\chip$ (Eq.~(3.4) in~\cite{chi_p}) and the effective spin-aligned parameter $\chieff$ (Eq.~(2) in~\cite{LIGOScientific:2020ibl}). 
Both parameters have been shown to be susceptible to biases due to the presence of glitches in real data~\cite{Payne:2022spz, Udall:2024ovp_191109}. 

\begin{table}[]
    \centering
    \begin{tblr}{colspec  = {lcl},
             rowsep=3pt,
             row{odd} = {bg=TableGrayDark},
             row{even} = {bg=TableGrayLight},
            }  
        \textbf{Parameter} & \textbf{Symbol} & \textbf{Prior} \\
        \hline
        Dimension & $D_g$ & U[1, 10] \\
        Wavelet central time & $t_0$ & U[-0.1, 0.1]\,s \\
        Wavelet central frequency & $f_0$ & U[16, 1024]\,Hz \\
        Wavelet quality factor & $Q$ & U[0.1, 40] \\
        Wavelet amplitude & $A$ & $\rho_i \in$ [1, 100] \\ 
        Wavelet phase & $\phi$ & U[0, $2\pi$] \\
    \end{tblr}
    \caption{Parameter definition, notation, and prior distribution for the glitch model, used to simulate glitches. 
    Here $\rho_i$ is the approximate SNR of the wavelet in Eq.~(12) in Ref.~\cite{Cornish:2020dwh}. This distribution is only used to simulate glitches. The wavelet central time distribution is centered at \{0, 0.25, 0.5, 1\}\,s after the center of the CBC time prior. }
    \label{tab:glitchPrior}
\end{table}

\subsubsection{Simulated glitches}\label{sec:glitch_model}

 We simulate glitches using $\BWCpp$'s glitch model for convenience, which consists of sums of sine-Gaussian Morlet-Gabor wavelets. 
 Such wavelets constitute an over-complete basis over a smooth function space and are thus flexible enough to mimic most noise transients in GW data. 
 In addition to each of the five parameters describing each wavelet (time, frequency, quality factor, amplitude, phase), the number of wavelets itself can be varied. 
 We adopt the $\BWCpp$ glitch prior as the distribution from which we draw glitches to simulate, listed in Table~\ref{tab:glitchPrior}. 
 
 The use of Morlet-Gabor wavelets allows us to quantify the glitch support in time and frequency space, e.g., Fig.~\ref{fig:schematic}. 
 Each wavelet is characterized by a central time $t_0$ and a central frequency $f_0$, around which its power decreases, forming a Gaussian envelope in both time and frequency. 
 In the time domain, the decay has an e-folding time scale of
\begin{equation}\label{eq:efold_time}
    \tau = \frac{Q}{2 \pi f_0}\,,
\end{equation}
 where $Q$ is the wavelet quality factor.
 Similarly, in the frequency domain, the decay e-folding frequency scale is $1 /\tau$. 
 So, a wavelet that is well-localized in time will be poorly localized in frequency and vice-versa.
 We use these estimates to approximately place the glitch with respect to the signal in subsequent sections.

 In reality, the majority of LIGO glitches cannot be described by a \emph{single} wavelet. Regardless, we expect our main conclusions to apply to more complex morphologies. Firstly, within $\BW$ glitches are described as a sum of wavelets. Studying the impact of a single wavelet therefore serves as a baseline. Secondly, the choice to use a single wavelet is driven by practical considerations: we can estimate its time-frequency footprint analytically with $f_0$ and Eq.~\eqref{eq:efold_time}. More generically, regardless of the exact glitch and signal morphologies, the amount of bias is determined by how much the glitch overlaps with the signal in time-frequency, and wavelets allow for a simple quantification of this.

\subsection{Glitch reweighting} \label{sec:reweighting}

We quantify the impact of glitches on the CBC posterior by reweighting the posterior from data \textit{without} to data \textit{with} a glitch. 
Equal-weighted samples from one (reference) probability distribution can be ``reweighted'' to approximate another (target) probability distribution; this is a form of importance sampling called ``reweighting". 
The exact formalism for reweighting is laid out in Sec.~II of ~\citet{Hourihane:2022ner_PTA}, which we adopt here. Reweighting provides a metric to compare how similar or disparate the two distributions are via the efficiency, 
\begin{equation}\label{eq:efficiency}
    \mathcal{E} = \frac{n_\mathrm{eff}}{N_s}\,,
\end{equation}
where $n_{\mathrm{eff}}$ is the  effective number of samples (Eq.~(11) in Ref.~\cite{Hourihane:2022ner_PTA}) after reweighting and $N_s$ is the original number of samples. 
Distributions with an efficiency of 1 are identical whereas disparate distributions have an efficiency of 0.

In this work we use reweighting in a somewhat novel way. 
While reweighting has been employed to change the likelihood function~\cite{Hourihane:2022ner_PTA}, the model~\cite{Payne:2019wmy, Romero-Shaw:2020thy, Romero-Shaw:2022xko, Romero-Shaw:2020aaj, Romero-Shaw:2021ual}, the prior, or to evaluate posteriors from a neural network approximate~\cite{Dax:2024mcn}, here we change the \textit{data}. 
That is, we use reweighting to transform a CBC posterior distribution in Gaussian noise, $p(\theta \mid \mathbf{d}_{\rm NG}) $, to a posterior distribution of identical data plus a glitch, $p(\theta \mid \mathbf{d}_{\rm G})$. 
The (unnormalized) log weights are 
 \begin{align}\label{eq:weights}
     \ln w(\theta) &=
       \ln\mathcal{L}(\mathbf{d}_{\rm G} | \theta) - \ln \mathcal{L}(\mathbf{d}_{\rm NG} | \theta) \nonumber\\
       &= -\frac{1}{2} |\mathbf{d}_{\rm NG} + \mathbf{g} - \mathbf{h}_\theta|^2 + \frac{1}{2}|\mathbf{d}_{\rm NG} - \mathbf{h}_\theta|^2\nonumber\\
       &= \left(\mathbf{g} | \mathbf{h}_\theta\right) - C_g\,,
 \end{align}
 where $C_g = (\mathbf{g} | \mathbf{g}) / 2  + (\mathbf{d} | \mathbf{g})$ is constant.

Reweighting has a number of advantages.
Firstly, it provides a sensitive, direct estimate of the difference between two probability distributions via the efficiency, Eq.~\eqref{eq:efficiency}.\footnote{Another option is to use the Kullback-Liebler divergence which can likewise be computed from the weights. However, efficiency is normalized so we prefer it.} 
Secondly, reweighting bypasses expensive stochastic sampling and allows us to consider a wide range of glitches from a single glitch-free set of samples. 
Thirdly, reweighting is not subject to sampling uncertainty when stochastically sampling from a posterior. 
That is, $\mathcal{E} \neq 1$ if and only if the standard deviation of the likelihood over the posterior changes between the approximate and the target distribution. 
Since glitches far away from the signal have a small impact on the posterior, we expect reweighting to be a particularly effective method for our purposes.

Such glitch-reweighting is possible on simulated data because we have access to glitch-free data. 
It is still possible to use glitch-reweighting on real data if there is a model for the glitch; for instance, the model used in glitch subtraction. 
The efficiency of such post-facto glitch-reweighting would serve as an estimate of the impact of that glitch on inference. 
In cases where the efficiency is low, it might be important to consider the uncertainty of the glitch model, e.g.~\cite{Hourihane:2022doe_glitch, Udall:2024ovp_191109}.  

\subsection{Jensen Shannon Divergence}\label{sec:JSD}

We also quantify the difference between two posteriors using the Jensen Shannon (JS) Divergence~\cite{Lin:1991}, see App.~A of ~\citet{LIGOScientific:2020ibl} for more details. 
The JS divergence between probability distributions $p$ and $q$, $D_{\rm JS}(p,q)$, is smoothed, normalized, and symmetrized. 
We adopt the threshold of $0.007\,$bit~\cite{LIGOScientific:2020ibl}.
For a Gaussian, this corresponds to a 20\% shift in the mean, which is larger than the 0.002\,bit expected between two identical posteriors from sampling uncertainty alone~\cite{Romero-Shaw:2020owr}. 
We compute the JS divergence of the 15 marginalized, 1-dimensional CBC posteriors and report the maximum
 \begin{equation}
     \max_\theta D^\theta_\mathrm{JS} = \max_{i\in{1...15}} D_\mathrm{JS}\left[ p(\theta_i | \mathbf{d}_{\rm G}), ~p(\theta_i | \mathbf{d}_{\rm NG})\right].
 \end{equation}

\section{Results}
\label{sec:results}

In this section we study how glitches impact CBC inference in each Region of Fig.~\ref{fig:schematic}.

\subsection{Region \rom{4}, Glitches above the signal}\label{sec:glitchesWayAboveSignal}

When the glitch and signal do not overlap in frequency, the CBC posterior is unaffected by the presence of the glitch, even when the glitch frequency is part of the analysis bandwidth.
The proof presented below is based on the fact that the noise-covariance is diagonal in the frequency domain, and thus the likelihood is a 1-dimensional integral in frequency.
Consider a GW signal only containing frequency content up to frequency $F$, coincident in time with a glitch that contains only frequency content above $F$. 
Then $(\mathbf{g} | \mathbf{h}) = 0$. 
If additionally the frequency content of $h_\theta(f > F) = 0 $ for \textit{all} $\theta$ in the CBC prior, $\pi$, it follows that, for all $\theta \in \pi$, the likelihood is independent of any glitch-signal cross-terms. 
That is: 
\begin{align}
\mathcal{L}\left(\mathbf{d} | \theta \right)&\propto
  \exp{\left [ -\frac{1}{2} \left(|\mathbf{d}_{\mathrm{NG}} + \mathbf{g}|^2 + |\mathbf{d}_{\mathrm{NG}} - \mathbf{h}_\theta|^2  - 2 (\mathbf{g} | \mathbf{h}_\theta) \right) \right]} \nonumber\\ 
&= \exp{\left [ -\frac{1}{2} \left(|\mathbf{d}_{\mathrm{NG}} + \mathbf{g}|^2 + |\mathbf{d}_{\mathrm{NG}} - \mathbf{h}_\theta|^2 \right) \right]}\,. 
\end{align}
The ${|\mathbf{d}_{\mathrm{NG}}+\mathbf{g}|^2}$ term is constant and does not depend on $\theta$.\footnote{In analyses where the glitch is also sampled, this term would no longer be constant.} 
Since the likelihood is not normalized in $\theta$, \textit{the extra glitch term does not change the shape of the posterior}. 
It will, however, scale the evidence, the expected value of the likelihood over the prior.


\subsection{Region \rom{3}, Glitches after the signal}\label{sec:glitchesAfterSignal}

\begin{figure*}
    \centering
    \includegraphics[width=\linewidth]{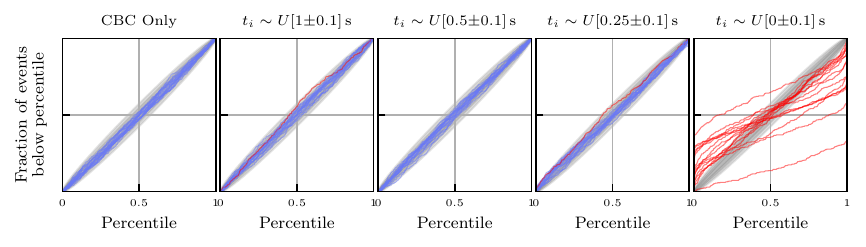}
    \caption{
    Percentile–Percentile (P-P) plots for various simulations, each drawn from the same CBC prior (Table~\ref{tab:CBCPrior}) but varying in glitch content (Table~\ref{tab:glitchPrior}). 
    The titles of each plot specify the relative time between the glitch distribution and the CBC time distribution, with the leftmost plot representing data without glitches and subsequent plots showing glitches progressively closer to the CBC. 
    Each plot comprises 400 simulations, with recovery performed using only a CBC model.
    Each plot includes 15 lines, one for each CBC parameter, displaying the cumulative distribution function of the percentiles of the true values within their marginal posteriors. 
    Lines are color-coded in red (blue) to indicate whether the parameter failed, $p \leq 0.05$ (passed, $p > 0.05$) the P-P test. 
    A failure rejects the null hypothesis: the percentiles of the true values are uniformly distributed across their posteriors. Three-sigma confidence intervals are plotted in gray. 
    Left: P–P plot when only the CBC model is simulated and recovered. All parameters pass the P-P test which serves as a baseline for the test and its implementation. 
    Center (left to right): P–P plots for glitches in Region \rom{3}. 
    Right: P–P plot for glitches in Region \rom{1}. When glitches overlap with the signals, all parameters fail the P-P test.\\
    }\label{fig:cbc_glitch_PP}
\end{figure*}

We now turn to glitches occurring after a GW event that do not overlap with the GW in time. 
The proof of Sec.~\ref{sec:glitchesWayAboveSignal}--glitches that do not overlap in frequency do not bias inference--no longer applies.
This is because the noise-covariance is no longer diagonal in the time domain, and distinct times are correlated, see Sec.~\ref{sec:whittle}.
Even if a glitch and signal do not overlap in time, the correlations between their times \textit{could} induce a bias. 
We study this bias for a population of signals and glitches in Sec.~\ref{sec:varying_distance}, and for individual glitches on a GW150914-like event in Secs.~\ref{sec:varying_SNR} and~\ref{sec:III_snr100_test}. 


\subsubsection{Percentile-Percentile Test}\label{sec:P-P}
 
We first introduce the percentile-percentile (P-P) test, a standard method to determine if an ensemble of posteriors are statistically robust~\cite{Gibbons2003, Cook_PP_test}.
In a P-P test, a set of simulations is generated by drawing parameters from the prior distribution of a model. 
The posterior is then computed with the same model and prior. 
If the prior reflects the underlying population (including the noise model) and the posteriors have the correct statistical coverage, the percentiles of the true parameter values within their respective posteriors should follow a uniform distribution. 
Consequently, the cumulative distribution function (CDF) of these percentiles should form a diagonal line with a slope of 1.

We perform this test with $400$ simulated CBC signals (without glitches) in Gaussian noise \and with parameters drawn from the CBC prior of Table~\ref{tab:CBCPrior}.
The leftmost panel of Fig.~\ref{fig:cbc_glitch_PP} shows the CDF for each of the 15 CBC model parameters. 
All parameters lie within the 3-sigma confidence region (indicated by the gray outline) as expected. 
This demonstrates that the $\BWCpp$ CBC sampler is unbiased under the conditions of the P-P test.

\subsubsection{Varying distance between glitch and signal}\label{sec:varying_distance} 

To assess the impact of glitches, we perform a variation of the standard P-P test described in Sec.~\ref{sec:P-P}. Instead of pure Gaussian noise, the simulated data now also contain glitches drawn from the distribution of Table~\ref{tab:glitchPrior}; their SNR follows their prior which peaks at SNR 5 and is bounded between 1 and 100.
The glitch time distribution is progressively moved closer to the signal. Still, just as in Sec.~\ref{sec:P-P}, the posterior is computed only over the CBC model, leaving the glitch unaccounted for. 
Five scenarios are compared in Fig.~\ref{fig:cbc_glitch_PP}: a control case with no glitches, and four cases with glitches whose time distributions have a width of $0.2\,$ and are centered at $1\,$s, $0.5$\,s, $0.25$\,s, and $0$\,s after the center of the CBC time prior. 
All simulations are performed with a single, LIGO Hanford (LHO) detector (to maximize the impact of glitches), and the Advanced LIGO sensitivity~\cite{LIGOScientific:2014pky}. 
By leaving the glitch unaccounted for during recovery and varying the timing of the glitch relative to the CBC, we explore how the timing of glitches impacts inference. 
 

Results are shown in Fig.~\ref{fig:cbc_glitch_PP}. 
The leftmost panel displays the control case with no glitches; all parameters pass the P–P test, indicating unbiased recovery. 
The center three panels (from left to right) depict results on data containing glitches after, but progressively closer to, the CBC signals. Recovery remains unaffected, with most of the 15 CBC parameters passing the test for each distance.
A few cases with $p<0.05$, are not unexpected given the large number of tests performed.
In contrast, the rightmost panel, corresponding to glitches coincident in time with the CBC signals, reveals that all parameters fail the P–P test, highlighting significant biases.
These result suggest that glitches occurring after a GW signal (Region~\rom{3}) have minimal impact on recovery for an ensemble of signals. 
However, glitches coincident with the signal (Region~\rom{1}) severely bias the recovered parameters, confirming previous results~\cite{Hourihane:2022doe_glitch, Ghonge:2023ksb, Pankow:2018qpo, Payne:2022spz_200129, Udall:2024ovp_191109, Macas:2022afm, Powell:2018csz, Kwok:2021zny, Mozzon:2021wam, Chatziioannou:2021ezd}.


\subsubsection{Can very loud glitches bias inference?}\label{sec:varying_SNR} 

\begin{table}[]
    \centering
    \begin{tblr}{
        colspec = {ll},
        rowsep = 3pt,
        width = \textwidth
    }
       \SetRow{bg=TableGrayDark}\textbf{Parameter} & \textbf{Value} \\
        \hline
        \SetRow{bg=TableGrayLight} Masses & $m_1 = 38.2\, M_\odot$, \\ 
        \SetRow{bg=TableGrayLight} & $m_2 = 32.9 \,M_\odot$  \\
        \SetRow{bg=TableGrayDark} Spin amplitudes & $\chi_1 = 0.998$, \\ 
        \SetRow{bg=TableGrayDark} & $\chi_2 = 0.126$  \\
        \SetRow{bg=TableGrayLight} Spin in-plane angle & $\phi_1 = 0.55\pi\,\mathrm{rad}$, \\ 
        \SetRow{bg=TableGrayLight} & $\phi_2=0.364 \pi\,\mathrm{rad}$  \\ 
        \SetRow{bg=TableGrayDark} Spin polar angle & $\theta_1=0.14\pi\,\mathrm{rad}$ \\ 
        \SetRow{bg=TableGrayDark} & $\theta_2=0.49\pi\,\mathrm{rad}$ \\ 
        \SetRow{bg=TableGrayLight} Hanford time & 1126259462.424\,s \\ 
        \SetRow{bg=TableGrayLight} & 2.42455\,s (in segment) \\
        \SetRow{bg=TableGrayDark} Luminosity distance & $D_L=415.66\,\mathrm{Mpc}$ \\
        \SetRow{bg=TableGrayLight} Inclination & $\iota = 0.88 \pi \,\mathrm{rad}$  \\ 
        \SetRow{bg=TableGrayDark} Right ascension & $\alpha = 0.35\pi\,\mathrm{rad}$ \\ 
        \SetRow{bg=TableGrayLight} Declination & $\delta = -0.36\pi\,\mathrm{rad}$ \\ 
        \SetRow{bg=TableGrayDark} Polarization & $\psi = 0.42\pi\,\mathrm{rad}$ \\
        \SetRow{bg=TableGrayLight} Coalescence phase & $\phi = 1.30\pi\,\mathrm{rad}$ \\ 
    \end{tblr}
    \caption{Simulated CBC parameters for Sec.~\ref{sec:varying_SNR}, Sec.~\ref{sec:III_snr100_test}, and Sec.~\ref{sec:snr_100_IIa}. }
    \label{tab:CBCInjection}
\end{table}

\begin{table}[]
    \centering
    \begin{tblr}{
        colspec = {ll},
        rowsep = 3pt,
    }          \SetRow{bg=TableGrayDark}\textbf{Parameter} & \textbf{Value} \\
    \hline
        \SetRow{bg=TableGrayLight} Glitch dimension & $D_g = 1$ \\
        \SetRow{bg=TableGrayDark} Wavelet central time & $t_1 = 2.38\,\mathrm{s}$ \\
        \SetRow{bg=TableGrayLight} Wavelet central frequency & $f_1 = 100\,\mathrm{Hz}$ \\
        \SetRow{bg=TableGrayDark} Wavelet quality factor & $Q_1 = 28.73$ \\
        \SetRow{bg=TableGrayLight} Wavelet amplitude & $A_1$: Varied such that \\
        \SetRow{bg=TableGrayLight} & $\rho_1 \in \{5, 10, 50, 100,$ \\
        \SetRow{bg=TableGrayLight} & $~~~~~~~~500, 1000, 5000\}$ \\
        \SetRow{bg=TableGrayDark} Wavelet phase & $\phi_1 = 1.61\pi\,\mathrm{rad}$ \\
    \end{tblr}
    \caption{Simulated glitch parameters used in Sec.~\ref{sec:varying_SNR}.}
    \label{tab:glitchInjectionVarySNR}
\end{table}

\begin{figure}
    \centering
    \includegraphics[
    ]{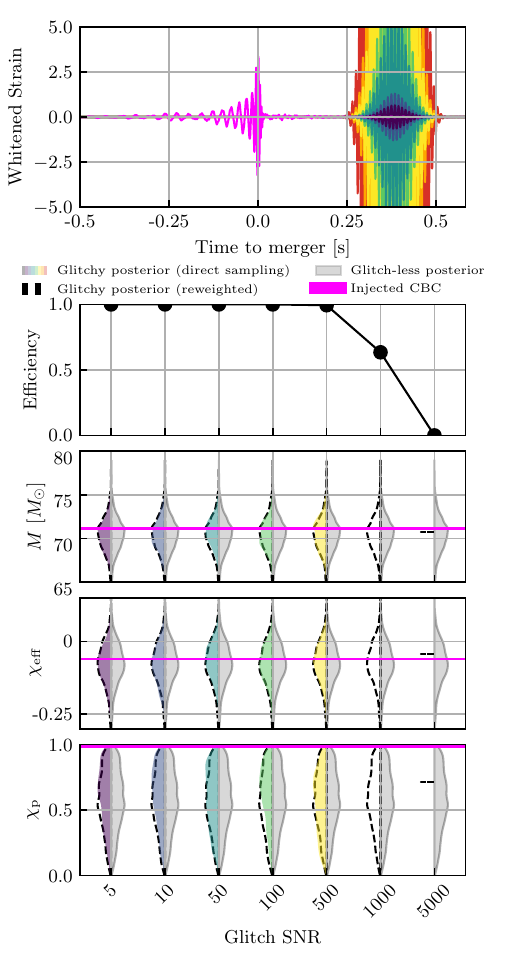}
    \caption{Top: Time domain whitened waveforms for the CBC (magenta) and a glitch with increasing SNR from 5 to 5000 (various colors) 0.38\,s after the signal, see Table~\ref{tab:CBCInjection} and Table~\ref{tab:glitchInjectionVarySNR} for details. Second down: Efficiency when reweighting from a posterior on data with no glitch to a posterior on data with a glitch as a function of the glitch SNR. 
    Bottom 3: 1-dimensional posteriors for select CBC parameters as a function of the glitch SNR. True values are marked in magenta. 
    The direct-sampled posterior on glitch-impacted data is colored and the corresponding reweighted posterior is marked with black dashed lines. To the right we show the control dataset in gray, a posterior recovered from data with identical Gaussian noise but without any glitch; all gray posteriors are identical. For glitch SNR $\geq 1000$ we omit the direct-sampled posteriors due to nonphysical waveform behavior, further discussed in App.~\ref{app:waveform_conditioning}. The efficiency, indicates that glitches louder than SNR 500 start impacting inference.}
    \label{fig:increasing_SNR}
\end{figure}

We now consider a single instance of a CBC and a glitch and study at which SNR a glitch can create a bias. 
We compare two datasets, again in a single LHO detector: 
\begin{enumerate}
    \item \textbf{Control Dataset:} A simulated GW150914-like signal with a network SNR of 25, injected into Gaussian noise. The CBC parameters are detailed in Table~\ref{tab:CBCInjection}.
    \item \textbf{Glitch-impacted Dataset:} The glitch-impacted dataset is identical to the control one in terms of the CBC and Gaussian noise realization but includes seven distinct glitch configurations.
    The glitches share all parameters, Table~\ref{tab:glitchInjectionVarySNR}, but SNR which we incrementally increase from 5 and 5000. 
\end{enumerate}
 
We analyze both datasets with $\BWCpp$, modeling \textit{only} the CBC signal, \textit{leaving again the glitch power entirely unmodeled}. 
Results for the control dataset serve as a baseline for comparison; if the glitches have no impact, the posteriors from the glitch-impacted dataset will match those from the glitch-free dataset. 
We compute glitch-impacted posteriors both with direct sampling ($\BWCpp$) and via reweighting from the control to the glitch-impacted dataset, see Sec.~\ref{sec:reweighting}. 
If the glitches have no impact, then the reweighting efficiency will be 1. 
  
We present the results in Fig.~\ref{fig:increasing_SNR}. 
The top panel shows the whitened time-domain waveforms of the CBC (magenta) and the glitches (various colors) for reference. 
Below that, we plot the reweighting efficiency and posteriors for select CBC parameters (total mass, $\chieff$, and $\chip$) as a function of glitch SNR. 
The direct-sampled, glitch-impacted posteriors are colored and the corresponding reweighted posteriors are plotted with black dashed lines. The fiducial, glitch-free posterior is displayed in gray. 
For glitch SNR $\leq 500$\footnote{
For reference, the glitch on the binary neutron star merger, GW170817, was SNR $\sim 800$~\cite{Pankow:2018qpo}.
 } we find that not only are the glitch-impacted posteriors \textit{visually} indistinguishable to the glitch-free one, but they are \textit{actually} identical; the reweighting efficiency between the two distributions is 100\%. 
 As the glitch SNR is increased, the efficiency does drop. 
At glitch SNR of 1000 the posteriors are still visually identical, but the efficiency is 80\%, suggesting it is a more sensitive test than visual comparison (it also depends on the full 15-dimensional posterior and not select marginal ones).
By a glitch SNR of 5000, the efficiency is zero, signaling severe biases.

\subsubsection{At what glitch SNR can we expect a bias?}~\label{sec:III_snr100_test}
\begin{figure*}
    \centering
    \includegraphics[]{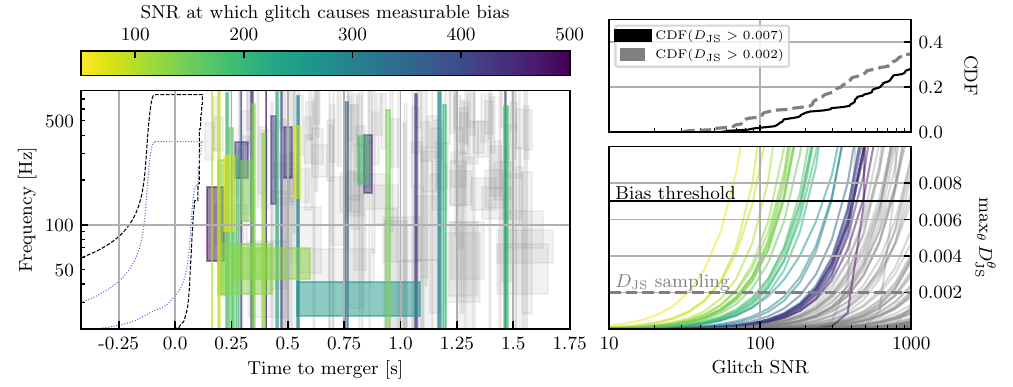}
    \caption{
    Left: Time-frequency locations of 200 glitches simulated in Region~\rom{3}; after the signal. The glitch time-frequency locations are colored by the SNR at which that glitch induces a measurable change in the CBC posterior and are gray if the requisite SNR is above 500. For reference, Region~\rom{1} (black-dashed) and the time-frequency content of the (2,2) mode across the CBC prior (blue dots) are displayed. 
    Bottom right: Maximum Jensen-Shannon divergence across CBC parameters between a CBC posterior on glitch-free data and the posterior on glitch-impacted data, plotted as a function of glitch SNR. Each curve corresponds to a glitch in the left figure. The black horizontal line is the threshold for posteriors considered distinct~\cite{LIGOScientific:2020ibl}. The gray dashed line is the JS divergence due only to stochastic sampling uncertainty~\cite{Romero-Shaw:2020owr}, plotted for reference. As the glitch SNR increases, so does does $D_\mathrm{JS}$. 
    Top right: Cumulative distribution function of the number of glitches that induce a requisite divergence as a function of glitch SNR. Glitches below SNR 50 never induce a measurable bias. Higher-SNR glitches might induce a bias if within $0.5\,$s from the signal merger. }
    \label{fig:regionIII_SNR_100}
\end{figure*}

\begin{table}[]
    \centering
    \begin{tblr}{
        colspec = {Q[l] Q[l]},
        rowsep = 3pt,
        colsep = 1pt,
        row{odd} = {bg=TableGrayDark},
        row{even} = {bg=TableGrayLight},
    }
        \textbf{Parameter} & \textbf{Value} \\
        \hline
        Glitch dimension & $D_g = 1$ \\
        Wavelet central time & $t_1 \sim U[0,4]\,$s \\
        Wavelet central frequency & $f_1 \sim U[20, 700]\,\Hz$ \\
        Wavelet quality factor & $Q_1 \sim U[0.1, 40]$ \\
        Wavelet amplitude & $A_1$: Varied \\
        Wavelet phase & $\phi_1 \sim U[0, 2\pi]$ \\
    \end{tblr}
    \caption{Parameter distribution of simulated glitches in Sec.~\ref{sec:III_snr100_test}, additionally restricted to Region~\rom{3}, and in Sec.~\ref{sec:snr_100_IIa}, likewise restricted to Region~\rom{2}a. }
    \label{tab:glitchInjectionIII&IIa}
\end{table}

Finally, we consider a random collection of 200 glitches per Table~\ref{tab:glitchInjectionIII&IIa} in Region~\rom{3} after the same CBC signal from Table~\ref{tab:CBCInjection}.
Since we expect glitch-induced biases to be low, we omit direct sampling, and instead only compute posteriors via reweighting. 
Since the weights are proportional to the glitch SNR, per Eq.~\eqref{eq:weights}, the reweighted posterior for a single glitch can be trivially scaled in SNR. 
We leverage this for each of the 200 glitches to compute the maximum JS divergence over all CBC parameters, and determine the glitch SNR at which it exceeds a JS threshold. 

Results are displayed in Fig.~\ref{fig:regionIII_SNR_100}. The time-frequency locations of the 200 simulated glitches are displayed in the leftmost panel, corresponding to three e-folds of exponential decay in each direction. 
Boxes are colored by the SNR at which that glitch induces a posterior with a JS divergence above 0.007 compared to glitch-free data.
Glitches are colored gray if the requisite SNR is above 500. 
On the bottom right we plot the maximum (among all 15 CBC parameters) JS divergence between the glitch-impacted and glitch-less datasets, plotted as a function of glitch SNR. 
Each curve corresponds to a glitch in the left figure, colored in the same manner. 
The top right shows the cumulative distribution of the number of glitches that induce an above-threshold divergence, plotted as a function of glitch SNR. 

The majority of glitches do not incur a bias unless their SNR is greater than $500$.
However, glitches closer to the signal (per the left plot, within 0.5\,s after merger) can incur a bias at SNR as low as $\sim56$.
Those are rare ($2\%$ at SNR 100 and $17\%$ at SNR 500) and will likely need mitigation. 
Even at SNR 1000, fewer than $40\%$ of the glitches induced a bias greater than one would expect from stochastic sampling.
Overall, \textit{no glitches below SNR 50 in Region~\rom{3} induce a measurable bias, no matter how close to (but confidently after) the signal}.


\begin{table}[]
    \centering
    \begin{tblr}{
        colspec = {Q[l] Q[l]},
        rowsep = 3pt,
        colsep = 1pt,
        row{odd} = {bg=TableGrayDark},
        row{even} = {bg=TableGrayLight},
    }
        \textbf{Parameter} & \textbf{Value} \\
        \hline
        Glitch dimension & $D_g = 1$ \\
        Wavelet central time & $t_1 = 2.04\,\mathrm{s}$ \\
        Wavelet central frequency & $f_1 \in \{25, 50, 100, 250, 500, 750\}\,\mathrm{Hz}$ \\
        Wavelet quality factor & $Q_1 = 28.73$ \\
        Wavelet amplitude & $A_1$: Varied such that $\rho_1 = 25$ \\
        Wavelet phase & $\phi_1 = 1.61\pi\,\mathrm{rad}$ \\
    \end{tblr}
    \caption{Simulated glitch parameters in Sec.~\ref{sec:region2A_1}.}
    \label{tab:glitchInjectionJustAbove}
\end{table}

\subsection{Region \rom{2}, Glitches sharing time and frequency content but not concurrently} \label{sec:glitchesJustAboveSignal}

\begin{figure*}
    \centering
    \includegraphics[width=\linewidth]{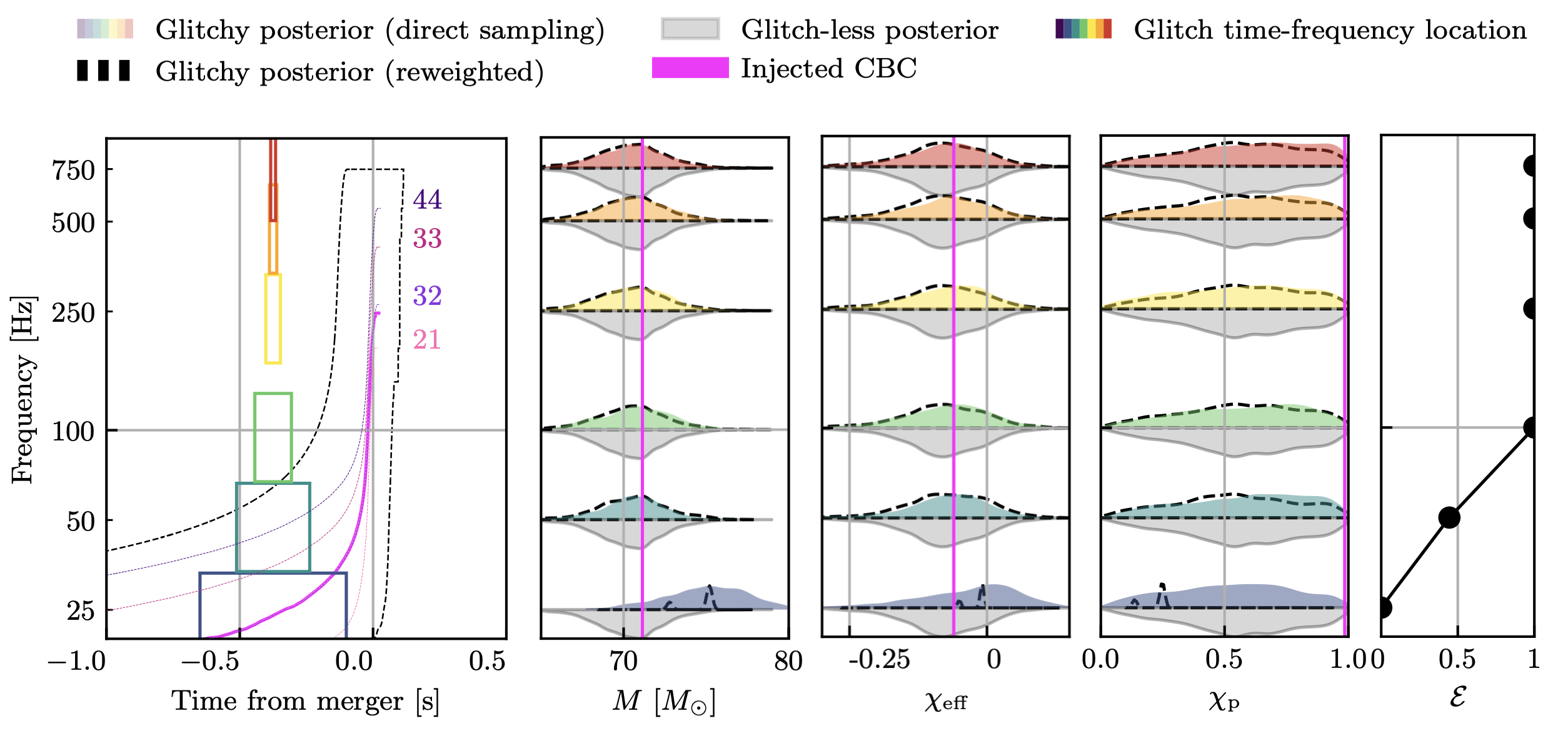}
    \caption{Inference results on a GW150914-like signal with SNR 25 glitches of increasing frequency. Left: Time-frequency track of the GW150914-like signal with glitches overlaid. Region~\rom{1} is outlined in black-dashed lines; it includes the time-frequency content of the entire CBC model prior. The solid magenta line shows the (2,2) frequency track of the simulated CBC and the dotted lines display the higher-order modes, each labeled with its corresponding color. The location in time-frequency space of the glitches is shown in a colored box, each displaying three e-folds of the exponential time and frequency glitch decay. The only glitches that share time-frequency content with the GW model are the ones at 25\,Hz and 50\,Hz. Center: 1-dimensional posteriors for select CBC parameters from data with a glitch at the corresponding frequency (y-axis and color). In the lower half of each violin plot we show the posterior from identical data but without a glitch (gray); all such posteriors are identical. On the top of each violin plot are the posteriors recovered in glitch-impacted data; the colored posterior are those recovered with $\BWCpp$ (direct sampling), and the black-dashed lines are the posterior obtained via reweighting. Right: Reweighting efficiency as a function of glitch frequency. Where the posteriors differ the most, at glitch frequency 25\,Hz, the efficiency drops to 0. At 50\,Hz, even though the posteriors are visually similar, the efficiency also dips to 50\%, meaning that the glitch is nonetheless impacting the posterior.}
    \label{fig:changingFreqSpectrogram}
\end{figure*}

Now we move to glitches that overlap with the GW signal (or its prior) in either time or frequency, but not concurrently. These are glitches ``just above," Region \rom{2}a, or ``just below," Region \rom{2}b, the signal. 
No pure time- or frequency-domain analysis can exclude those with analysis window redefinition. 

\subsubsection{Region \rom{2}a: Glitches just above the signal}
\label{sec:region2A_1}

We follow the methodology of Sec.~\ref{sec:varying_SNR} and consider two simulated datasets:

\begin{enumerate}
    \item \textbf{Control Dataset:} identical to that in Sec~\ref{sec:varying_SNR}, with a GW150914-like signal given in Table~\ref{tab:CBCInjection} in Gaussian noise. 
    \item \textbf{Glitch-impacted Dataset:} identical to the control dataset in terms of CBC and Gaussin noise, but it includes six distinct glitch configurations. 
    All glitches have an SNR of 25 and are centered ${\sim}13$ cycles before merger; further parameters are given in Table~\ref{tab:glitchInjectionJustAbove}. The frequency of the glitch is incrementally increased, starting with one that overlaps with the CBC's (2,2) mode and continuing until the glitch is outside the entire CBC model prior.
\end{enumerate}

The left panel of Fig.~\ref{fig:changingFreqSpectrogram} displays the time-frequency tracks for the modes of the injected CBC signal (colored lines), overlaid with the time-frequency support of the simulated glitches (colored boxes).
The solid magenta line represents the dominant (2,2) frequency track, while the dashed lines indicate higher-order modes. 
The dashed black line encloses the CBC model prior, including all higher-order modes.
The glitches at 25\,Hz and 50\,Hz overlap with the time-frequency GW prior, but only the glitch at 25\,Hz overlaps with the (2,2) mode. 

As always, we analyze the data leaving the glitch unmodeled.
In the central panels of Fig.~\ref{fig:changingFreqSpectrogram} we plot select marginalized CBC posteriors corresponding to each glitch (same color as the boxes) as well as the fiducial, glitch-less posterior (gray). 
The y-axis corresponds to the glitch central frequency. 
The black, dashed lines are again the glitch-reweighted distributions which should be identical to the colored distributions (for sufficiently high efficiency). 
Above 25\,Hz, all distributions agree on visual inspection for all parameters.
Glitches that do not overlap with the signal (2,2) mode do not have a noticeable visual effect on the posteriors. 

The 25\,Hz glitch overlaps the CBC's (2,2) mode in time-frequency space, and thus falls squarely into Region~\rom{1}. It is clearly biasing inference, increasing the total mass and $\chieff$. 
The reweighing efficiency (right panel) is close to 0\%, and thus reweighting is unable to reconstruct the direct-sampled posteriors. 

Glitches with central frequencies 100\,Hz and above have efficiencies close to 100\%, meaning that posteriors from data with and without the glitch are indistinguishable.
This is visually evident for $M, \chieff$, and $ \chip$ (middle panels), but the efficiency is a stronger test that considers the full 15-dimensional posteriors.
Thus, glitches that do not overlap in time-frequency with the CBC model prior do not impact inference in this example.

The 50\,Hz glitch presents a middle case: identical-looking posteriors but an efficiency of 50\%, indicating a measurable bias.
 This is because that glitch shares time-frequency content with the signal's high-order modes as well as the general CBC model prior (left panel). 
Glitches that overlap with a signal's high-order modes can therefore still lead to small biases, even if the dominant (2,2) mode is not impacted. 

\subsubsection{At what glitch SNR can we expect a bias?}~\label{sec:snr_100_IIa}

\begin{figure*}
    \centering
    \includegraphics[]{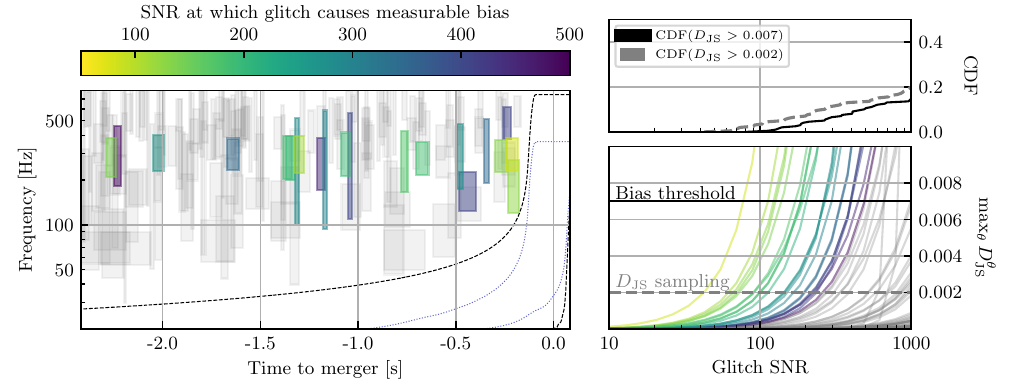}
    \caption{Left: Time-frequency locations of 200 glitches simulated in Region~\rom{2}a. The glitch time-frequency locations are colored by the SNR at which that glitch induces a measurable change in the CBC posterior and are gray if the requisite SNR is above 500. For reference, Region~\rom{1} (black-dashed) and the time-frequency content of the (2,2) mode across the CBC prior (blue dots) are displayed. 
    Bottom right: Maximum Jensen-Shannon divergence across CBC parameters between a CBC posterior on glitch-free data and the posterior on glitch-impacted data, plotted as a function of glitch SNR. Each curve corresponds to a glitch in the left figure, and are colored in the same manner. The black horizontal line is the threshold for posteriors considered distinct~\cite{LIGOScientific:2020ibl}. The gray dashed line is the JS divergence expected to arise due to stochastic sampling uncertainty~\cite{Romero-Shaw:2020owr}, displayed for reference. As the glitch SNR increases, so does $D_\mathrm{JS}$. 
    Top right: Cumulative distribution function of the number of glitches that induced a requisite divergence as a function of glitch SNR.
Only a single glitch with SNR less than 100 induces a measurable bias. Even at SNR 1000 fewer than $20\%$ of the glitches induce a difference greater than one would expect from stochastic sampling.}
    \label{fig:region_IIa_snr100}
\end{figure*}

 Finally, we repeat the analysis of Sec.~\ref{sec:III_snr100_test} with 200 random glitches in Region~\rom{2}a. 
 The analysis setup is identical and we refer to that discussion for details.
We present our results in Fig.~\ref{fig:region_IIa_snr100} in similar format to Fig.~\ref{fig:regionIII_SNR_100}. 
All glitches that result in a bias for SNR less than 500 have central frequencies close to the CBC merger frequency, 150\,Hz to 400\,Hz.
Even then, biases are rare.
\textit{Only a single glitch} (0.005\%) causes a bias by SNR 100. 
Somewhat surprisingly, glitches in this region have, on average, a smaller impact on inference than those in Region~\rom{3}.

\subsubsection{Region \rom{2}b: Glitches just below the signal}

Since we are focusing on high-mass events (where inference is most heavily impacted by glitches~\cite{Hourihane:2022doe_glitch}), Region \rom{2}b is very small. 
Figure~\ref{fig:schematic} shows that there is very little time-frequency space below the waveform and outside the prior.
We therefore do not perform any analyses here, instead expecting similar results to Region~\rom{2}a. 
Region \rom{2}b will be bigger (and could contain glitches) for lower-mass signals.


\subsection{Region \rom{1}, Glitches on top of the signal}\label{sec:glitchesOnTopOfSignal}

It has been shown both in real events~\cite{Payne:2022spz_200129, Udall:2024ovp_191109, Pankow:2018qpo} and simulations~\cite{Hourihane:2022doe_glitch} that glitches overlapping GWs can be detrimental to inference. 
We do not present further dedicated studies of this Region here, but nonetheless confirm the previous results in the right panel on Fig.~\ref{fig:cbc_glitch_PP} and Fig.~\ref{fig:changingFreqSpectrogram}.
The former shows that a population of glitches that overlap the time-frequency prior of a population of CBCs will results in strong biases.
The ``s-curves" in a P-P plot are indicative of the standard deviation of the posterior being incorrect.  
The latter shows that glitches that overlap with dominant signal mode can lead to large (visible) biases, but a smaller impact is also expected when the glitch overlaps the higher-order modes.

\section{Conclusion}\label{sec:conclusion}

The purpose of this study was to determine if there are time-frequency locations where one can ignore glitches nearby a transient GW, even when it is in the analysis window. 
 We split our findings and recommendations into the same Regions as in Fig.~\ref{fig:schematic}. 

 \begin{itemize}
     \item Region \rom{1} includes glitches that are coincident in time-frequency with the GW event and have been shown to impact inference on real signals~\cite{Payne:2022spz, Udall:2024ovp_191109, Pankow:2018qpo}. We confirm these results even at the population level; such glitches need to be carefully mitigated. 
     \item  Region \rom{2} includes glitches just above and just below the GW model prior in time-frequency. That is, these glitches overlap in time and in frequency with the signal, but not concurrently. Glitches in this region do not impacting inference and thus do not require mitigation, unless they have SNR above 50 \textit{and} are close in frequency to the CBC merger frequency. 
     \item  Region \rom{3} includes glitches that share frequency content with the CBC and its prior, but occur \textit{after} the GW, and thus do not overlap in time. For a fiducial population of glitches, biases are not expected for a glitch SNR below 50, or an SNR below 100 if the glitch is more than 0.5\,s after the merger. 
Mitigation is again not strictly required, though in that case it is significantly simpler to perform though inpainting~\cite{Zackay:2019kkv}, gating~\cite{Zweizig:gating}, or modeling solely the glitch with $\BWCpp$ with an appropriate analysis window that excludes the CBC~\cite{Davis:2022ird}.  
     \item  Region \rom{4} includes glitches that do not share frequency content with the GW signal. Even if these frequencies are included in the analysis window, they cannot impact inference. 
 \end{itemize}
 
Throughout this study we have assumed that the noise PSD is perfectly known and can be estimated regardless of the glitch. 
When estimating the PSD ``on source''~\cite{Littenberg:2014oda, Cornish:2020dwh,Chatziioannou:2019} the glitch model is present to account for non Gaussian noise that could affect the PSD is the glitch is loud. 
The uncertainty associated with the power spectrum estimate is also not accounted for here, although such uncertainty is a subdominant effect in CBC analyses on \textit{glitch-subtracted} data~\cite{Plunkett:2022zmx}. 
Other noise-mitigation techniques such as inpainting do require knowledge of the PSD which could be estimated off-source.


 In summary, glitches outside Region~\rom{1} in Fig~\ref{fig:schematic} are unlikely to affect parameter estimation unless they are sufficiently loud and close to the merger in time and/or frequency.
Though the results quantifying the glitch SNR required for biases were based on a single, representative CBC signal, we probed a large number of potential glitch parameters and time-frequency configurations.
Moreover, the population analysis of Sec.~\ref{sec:P-P} confirms these expectations for CBC signals drawn from the prior.  
 Quieter glitches (conservatively, below SNR 50) can thus be left in the data. 
 Loud glitches are fortunately much rarer than their low SNR counterparts.

\section{Acknowledgments}\label{sec:acknowledgments}

Thank you to Colm Talbot, Reed Essick, Jacob Golomb, Lucy Thomas, and Ryan Magee for useful discussions about the $\texttt{IMRPhenomXPHM}$ waveforms systematics. Thank you to Eliot Finch for help with determining the time-frequency location of the waveform ringdown. Thank you to Derek Davis for their expertise on glitches and glitch mitigation. Thank you to Jane Glanzer and Siddharth Soni for help with glitch rate estimates as a function of SNR. 
 SH was supported by the National Science Foundation Graduate Research Fellowship under Grant DGE-1745301. 
KC was supported by National Science Foundation Grant PHY-2409001.
 This material is based upon work supported by NSF’s LIGO Laboratory which is a major facility fully funded by the National Science Foundation.
 LIGO was constructed by the California Institute of Technology and Massachusetts Institute of Technology with funding from the National Science Foundation, and operates under cooperative agreement PHY-2309200. 
 The authors are grateful for computational resources provided by the LIGO Laboratory and supported by National Science Foundation Grants PHY-0757058 and PHY-0823459.
 This work made use of $\BWCpp$~\cite{bayeswave_cpp}, \texttt{gwpy}~\cite{duncan_macleod_2020_3598469}, \texttt{scipy}~\cite{2020SciPy-NMeth}, \texttt{matplotlib}~\cite{Hunter:2007}, \texttt{pandas}~\cite{reback2020pandas}, and \texttt{numpy}~\cite{harris2020array}.

\appendix

\section{Waveform Conditioning and Systematics}
\label{app:waveform_conditioning}

\begin{figure}
    \centering
    \includegraphics[]{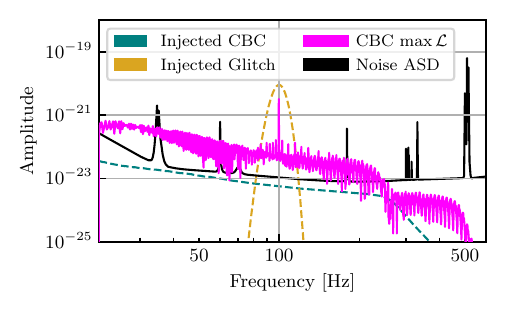}
    \caption{Spectral amplitude for the SNR 5000 glitch from Sec.~\ref{sec:varying_SNR}. We show the noise ASD (square root of the PSD) in black as well as the simulated CBC (gold) and glitch (green). The pink line shows the maximum likelihood posterior sample recovered when directly sampling the posterior with $\BWCpp$ and $\texttt{IMRPhenomXPHM}$. The sampler stumbled upon a rare waveform pathology that resulted in a spike in frequency that was able to partially fit the glitch. }
    \label{fig:waveform_systematic}
\end{figure}

In Sec.~\ref{sec:varying_SNR}, while we showed that the reweighting efficiency decreases for glitch SNRs of 1000 and 5000, we omitted posteriors created by direct-sampling on the glitch-impacted data. 
This was because in these cases the CBC model ended up fitting part of the glitch due to some waveform unphysical behavior. 
In Fig.~\ref{fig:waveform_systematic}, we show the maximum likelihood waveform recovered on that data, which has a clear, non-physical spike caused by a known failure in the multibanding~\cite{Pratten:2020ceb} of the waveform.

Regardless of waveform systematics, we recover a drop in efficiency due to the presence of these high SNR glitches. This efficiency drop (computed via reweighting) \textit{cannot} be caused by the aforementioned waveform systematic. On glitch-free data, such a pathological waveform is so disfavored by the likelihood and occupies a minute region of parameter space, that it has a vanishing probability of appearing in a direct-sampled posterior.
We interpret the fact that the $\BWCpp$ sampler located this pathological behavior as a testament to its efficacy. 

If such inference was obtained from a real signal, the issue would be immediately obvious. Nonetheless, we recommend that glitches with an SNR above 500 be mitigated.

\bibliographystyle{apsrev4-2} 
\bibliography{all_refs.bib}

\begin{thebibliography}{59}%
\makeatletter
\providecommand \@ifxundefined [1]{%
 \@ifx{#1\undefined}
}%
\providecommand \@ifnum [1]{%
 \ifnum #1\expandafter \@firstoftwo
 \else \expandafter \@secondoftwo
 \fi
}%
\providecommand \@ifx [1]{%
 \ifx #1\expandafter \@firstoftwo
 \else \expandafter \@secondoftwo
 \fi
}%
\providecommand \natexlab [1]{#1}%
\providecommand \enquote  [1]{``#1''}%
\providecommand \bibnamefont  [1]{#1}%
\providecommand \bibfnamefont [1]{#1}%
\providecommand \citenamefont [1]{#1}%
\providecommand \href@noop [0]{\@secondoftwo}%
\providecommand \href [0]{\begingroup \@sanitize@url \@href}%
\providecommand \@href[1]{\@@startlink{#1}\@@href}%
\providecommand \@@href[1]{\endgroup#1\@@endlink}%
\providecommand \@sanitize@url [0]{\catcode `\\12\catcode `\$12\catcode `\&12\catcode `\#12\catcode `\^12\catcode `\_12\catcode `\%12\relax}%
\providecommand \@@startlink[1]{}%
\providecommand \@@endlink[0]{}%
\providecommand \url  [0]{\begingroup\@sanitize@url \@url }%
\providecommand \@url [1]{\endgroup\@href {#1}{\urlprefix }}%
\providecommand \urlprefix  [0]{URL }%
\providecommand \Eprint [0]{\href }%
\providecommand \doibase [0]{https://doi.org/}%
\providecommand \selectlanguage [0]{\@gobble}%
\providecommand \bibinfo  [0]{\@secondoftwo}%
\providecommand \bibfield  [0]{\@secondoftwo}%
\providecommand \translation [1]{[#1]}%
\providecommand \BibitemOpen [0]{}%
\providecommand \bibitemStop [0]{}%
\providecommand \bibitemNoStop [0]{.\EOS\space}%
\providecommand \EOS [0]{\spacefactor3000\relax}%
\providecommand \BibitemShut  [1]{\csname bibitem#1\endcsname}%
\let\auto@bib@innerbib\@empty
\bibitem [{\citenamefont {Abbott}\ \emph {et~al.}(2023{\natexlab{a}})\citenamefont {Abbott} \emph {et~al.}}]{gwtc3}%
  \BibitemOpen
  \bibfield  {author} {\bibinfo {author} {\bibfnamefont {R.}~\bibnamefont {Abbott}} \emph {et~al.} (\bibinfo {collaboration} {KAGRA, VIRGO, LIGO Scientific}),\ }\href {https://doi.org/10.1103/PhysRevX.13.041039} {\bibfield  {journal} {\bibinfo  {journal} {Phys. Rev. X}\ }\textbf {\bibinfo {volume} {13}},\ \bibinfo {pages} {041039} (\bibinfo {year} {2023}{\natexlab{a}})},\ \Eprint {https://arxiv.org/abs/2111.03606} {arXiv:2111.03606 [gr-qc]} \BibitemShut {NoStop}%
\bibitem [{\citenamefont {Abbott}\ \emph {et~al.}(2021{\natexlab{a}})\citenamefont {Abbott} \emph {et~al.}}]{O3a_catalogue}%
  \BibitemOpen
  \bibfield  {author} {\bibinfo {author} {\bibfnamefont {R.}~\bibnamefont {Abbott}} \emph {et~al.} (\bibinfo {collaboration} {LIGO Scientific, Virgo}),\ }\href {https://doi.org/10.1103/PhysRevX.11.021053} {\bibfield  {journal} {\bibinfo  {journal} {Phys. Rev. X}\ }\textbf {\bibinfo {volume} {11}},\ \bibinfo {pages} {021053} (\bibinfo {year} {2021}{\natexlab{a}})},\ \Eprint {https://arxiv.org/abs/2010.14527} {arXiv:2010.14527 [gr-qc]} \BibitemShut {NoStop}%
\bibitem [{\citenamefont {Abbott}\ \emph {et~al.}(2023{\natexlab{b}})\citenamefont {Abbott} \emph {et~al.}}]{KAGRA:2021duu}%
  \BibitemOpen
  \bibfield  {author} {\bibinfo {author} {\bibfnamefont {R.}~\bibnamefont {Abbott}} \emph {et~al.} (\bibinfo {collaboration} {KAGRA, VIRGO, LIGO Scientific}),\ }\href {https://doi.org/10.1103/PhysRevX.13.011048} {\bibfield  {journal} {\bibinfo  {journal} {Phys. Rev. X}\ }\textbf {\bibinfo {volume} {13}},\ \bibinfo {pages} {011048} (\bibinfo {year} {2023}{\natexlab{b}})},\ \Eprint {https://arxiv.org/abs/2111.03634} {arXiv:2111.03634 [astro-ph.HE]} \BibitemShut {NoStop}%
\bibitem [{\citenamefont {Abbott}\ \emph {et~al.}(2021{\natexlab{b}})\citenamefont {Abbott} \emph {et~al.}}]{LIGOScientific:2021sio}%
  \BibitemOpen
  \bibfield  {author} {\bibinfo {author} {\bibfnamefont {R.}~\bibnamefont {Abbott}} \emph {et~al.} (\bibinfo {collaboration} {LIGO Scientific, VIRGO, KAGRA}),\ }\href {https://doi.org/10.48550/arXiv.2112.06861} {\bibfield  {journal} {\bibinfo  {journal} {arXiv e-prints}\ ,\ \bibinfo {eid} {arXiv:2112.06861}} (\bibinfo {year} {2021}{\natexlab{b}})},\ \Eprint {https://arxiv.org/abs/2112.06861} {arXiv:2112.06861 [gr-qc]} \BibitemShut {NoStop}%
\bibitem [{\citenamefont {Abbott}\ \emph {et~al.}(2016)\citenamefont {Abbott} \emph {et~al.}}]{GW150914}%
  \BibitemOpen
  \bibfield  {author} {\bibinfo {author} {\bibfnamefont {B.}~\bibnamefont {Abbott}} \emph {et~al.} (\bibinfo {collaboration} {LIGO Scientific, Virgo}),\ }\href {https://doi.org/10.1103/PhysRevLett.116.061102} {\bibfield  {journal} {\bibinfo  {journal} {Phys. Rev. Lett.}\ }\textbf {\bibinfo {volume} {116}},\ \bibinfo {pages} {061102} (\bibinfo {year} {2016})},\ \Eprint {https://arxiv.org/abs/1602.03837} {arXiv:1602.03837 [gr-qc]} \BibitemShut {NoStop}%
\bibitem [{\citenamefont {Aasi}\ \emph {et~al.}(2015)\citenamefont {Aasi} \emph {et~al.}}]{LIGOScientific:2014pky}%
  \BibitemOpen
  \bibfield  {author} {\bibinfo {author} {\bibfnamefont {J.}~\bibnamefont {Aasi}} \emph {et~al.} (\bibinfo {collaboration} {LIGO Scientific}),\ }\href {https://doi.org/10.1088/0264-9381/32/7/074001} {\bibfield  {journal} {\bibinfo  {journal} {Class. Quant. Grav.}\ }\textbf {\bibinfo {volume} {32}},\ \bibinfo {pages} {074001} (\bibinfo {year} {2015})},\ \Eprint {https://arxiv.org/abs/1411.4547} {arXiv:1411.4547 [gr-qc]} \BibitemShut {NoStop}%
\bibitem [{\citenamefont {Acernese}\ \emph {et~al.}(2015)\citenamefont {Acernese} \emph {et~al.}}]{VIRGO:2014yos}%
  \BibitemOpen
  \bibfield  {author} {\bibinfo {author} {\bibfnamefont {F.}~\bibnamefont {Acernese}} \emph {et~al.} (\bibinfo {collaboration} {VIRGO}),\ }\href {https://doi.org/10.1088/0264-9381/32/2/024001} {\bibfield  {journal} {\bibinfo  {journal} {Class. Quant. Grav.}\ }\textbf {\bibinfo {volume} {32}},\ \bibinfo {pages} {024001} (\bibinfo {year} {2015})},\ \Eprint {https://arxiv.org/abs/1408.3978} {arXiv:1408.3978 [gr-qc]} \BibitemShut {NoStop}%
\bibitem [{\citenamefont {Collaboration}\ and\ \citenamefont {Collaboration}(2024)}]{LIGO_Public_Alerts}%
  \BibitemOpen
  \bibfield  {author} {\bibinfo {author} {\bibfnamefont {L.~S.}\ \bibnamefont {Collaboration}}\ and\ \bibinfo {author} {\bibfnamefont {V.}~\bibnamefont {Collaboration}},\ }\href@noop {} {\bibinfo {title} {Ligo/virgo public alerts (o4)}},\ \bibinfo {howpublished} {\url{https://gracedb.ligo.org/superevents/public/O4/}} (\bibinfo {year} {2024}),\ \bibinfo {note} {accessed: 2024-10-09}\BibitemShut {NoStop}%
\bibitem [{\citenamefont {Soni}\ \emph {et~al.}(2025)\citenamefont {Soni}, \citenamefont {Berger}, \citenamefont {Davis} \emph {et~al.}}]{LIGO:2024kkz_detcharO4}%
  \BibitemOpen
  \bibfield  {author} {\bibinfo {author} {\bibfnamefont {S.}~\bibnamefont {Soni}}, \bibinfo {author} {\bibfnamefont {B.~K.}\ \bibnamefont {Berger}}, \bibinfo {author} {\bibfnamefont {D.}~\bibnamefont {Davis}}, \emph {et~al.},\ }\href {https://doi.org/10.1088/1361-6382/adc4b6} {\bibfield  {journal} {\bibinfo  {journal} {Classical and Quantum Gravity}\ }\textbf {\bibinfo {volume} {42}},\ \bibinfo {pages} {085016} (\bibinfo {year} {2025})}\BibitemShut {NoStop}%
\bibitem [{\citenamefont {Hourihane}\ \emph {et~al.}(2022)\citenamefont {Hourihane}, \citenamefont {Chatziioannou}, \citenamefont {Wijngaarden}, \citenamefont {Davis}, \citenamefont {Littenberg},\ and\ \citenamefont {Cornish}}]{Hourihane:2022doe_glitch}%
  \BibitemOpen
  \bibfield  {author} {\bibinfo {author} {\bibfnamefont {S.}~\bibnamefont {Hourihane}}, \bibinfo {author} {\bibfnamefont {K.}~\bibnamefont {Chatziioannou}}, \bibinfo {author} {\bibfnamefont {M.}~\bibnamefont {Wijngaarden}}, \bibinfo {author} {\bibfnamefont {D.}~\bibnamefont {Davis}}, \bibinfo {author} {\bibfnamefont {T.}~\bibnamefont {Littenberg}},\ and\ \bibinfo {author} {\bibfnamefont {N.}~\bibnamefont {Cornish}},\ }\href {https://doi.org/10.1103/PhysRevD.106.042006} {\bibfield  {journal} {\bibinfo  {journal} {Phys. Rev. D}\ }\textbf {\bibinfo {volume} {106}},\ \bibinfo {pages} {042006} (\bibinfo {year} {2022})},\ \Eprint {https://arxiv.org/abs/2205.13580} {arXiv:2205.13580 [gr-qc]} \BibitemShut {NoStop}%
\bibitem [{\citenamefont {Ghonge}\ \emph {et~al.}(2023)\citenamefont {Ghonge}, \citenamefont {Brandt}, \citenamefont {Sullivan}, \citenamefont {Millhouse}, \citenamefont {Chatziioannou}, \citenamefont {Clark}, \citenamefont {Littenberg}, \citenamefont {Cornish}, \citenamefont {Hourihane},\ and\ \citenamefont {Cadonati}}]{Ghonge:2023ksb}%
  \BibitemOpen
  \bibfield  {author} {\bibinfo {author} {\bibfnamefont {S.}~\bibnamefont {Ghonge}}, \bibinfo {author} {\bibfnamefont {J.}~\bibnamefont {Brandt}}, \bibinfo {author} {\bibfnamefont {J.~M.}\ \bibnamefont {Sullivan}}, \bibinfo {author} {\bibfnamefont {M.}~\bibnamefont {Millhouse}}, \bibinfo {author} {\bibfnamefont {K.}~\bibnamefont {Chatziioannou}}, \bibinfo {author} {\bibfnamefont {J.~A.}\ \bibnamefont {Clark}}, \bibinfo {author} {\bibfnamefont {T.}~\bibnamefont {Littenberg}}, \bibinfo {author} {\bibfnamefont {N.}~\bibnamefont {Cornish}}, \bibinfo {author} {\bibfnamefont {S.}~\bibnamefont {Hourihane}},\ and\ \bibinfo {author} {\bibfnamefont {L.}~\bibnamefont {Cadonati}},\ }\href@noop {} {\bibfield  {journal} {\bibinfo  {journal} {Accepted: Phys. Rev. D}\ } (\bibinfo {year} {2023})},\ \Eprint {https://arxiv.org/abs/2311.09159} {arXiv:2311.09159 [gr-qc]} \BibitemShut {NoStop}%
\bibitem [{\citenamefont {Pankow}\ \emph {et~al.}(2018)\citenamefont {Pankow} \emph {et~al.}}]{Pankow:2018qpo}%
  \BibitemOpen
  \bibfield  {author} {\bibinfo {author} {\bibfnamefont {C.}~\bibnamefont {Pankow}} \emph {et~al.},\ }\href {https://doi.org/10.1103/PhysRevD.98.084016} {\bibfield  {journal} {\bibinfo  {journal} {Phys. Rev.}\ }\textbf {\bibinfo {volume} {D98}},\ \bibinfo {pages} {084016} (\bibinfo {year} {2018})},\ \Eprint {https://arxiv.org/abs/1808.03619} {arXiv:1808.03619 [gr-qc]} \BibitemShut {NoStop}%
\bibitem [{\citenamefont {Payne}\ \emph {et~al.}(2022{\natexlab{a}})\citenamefont {Payne}, \citenamefont {Hourihane}, \citenamefont {Golomb}, \citenamefont {Udall}, \citenamefont {Udall}, \citenamefont {Davis},\ and\ \citenamefont {Chatziioannou}}]{Payne:2022spz_200129}%
  \BibitemOpen
  \bibfield  {author} {\bibinfo {author} {\bibfnamefont {E.}~\bibnamefont {Payne}}, \bibinfo {author} {\bibfnamefont {S.}~\bibnamefont {Hourihane}}, \bibinfo {author} {\bibfnamefont {J.}~\bibnamefont {Golomb}}, \bibinfo {author} {\bibfnamefont {R.}~\bibnamefont {Udall}}, \bibinfo {author} {\bibfnamefont {R.}~\bibnamefont {Udall}}, \bibinfo {author} {\bibfnamefont {D.}~\bibnamefont {Davis}},\ and\ \bibinfo {author} {\bibfnamefont {K.}~\bibnamefont {Chatziioannou}},\ }\href {https://doi.org/10.1103/PhysRevD.106.104017} {\bibfield  {journal} {\bibinfo  {journal} {Phys. Rev. D}\ }\textbf {\bibinfo {volume} {106}},\ \bibinfo {pages} {104017} (\bibinfo {year} {2022}{\natexlab{a}})},\ \Eprint {https://arxiv.org/abs/2206.11932} {arXiv:2206.11932 [gr-qc]} \BibitemShut {NoStop}%
\bibitem [{\citenamefont {Udall}\ \emph {et~al.}(2025)\citenamefont {Udall}, \citenamefont {Hourihane}, \citenamefont {Miller}, \citenamefont {Davis}, \citenamefont {Chatziioannou}, \citenamefont {Isi},\ and\ \citenamefont {Deshong}}]{Udall:2024ovp_191109}%
  \BibitemOpen
  \bibfield  {author} {\bibinfo {author} {\bibfnamefont {R.}~\bibnamefont {Udall}}, \bibinfo {author} {\bibfnamefont {S.}~\bibnamefont {Hourihane}}, \bibinfo {author} {\bibfnamefont {S.}~\bibnamefont {Miller}}, \bibinfo {author} {\bibfnamefont {D.}~\bibnamefont {Davis}}, \bibinfo {author} {\bibfnamefont {K.}~\bibnamefont {Chatziioannou}}, \bibinfo {author} {\bibfnamefont {M.}~\bibnamefont {Isi}},\ and\ \bibinfo {author} {\bibfnamefont {H.}~\bibnamefont {Deshong}},\ }\href {https://doi.org/10.1103/PhysRevD.111.024046} {\bibfield  {journal} {\bibinfo  {journal} {Phys. Rev. D}\ }\textbf {\bibinfo {volume} {111}},\ \bibinfo {pages} {024046} (\bibinfo {year} {2025})},\ \Eprint {https://arxiv.org/abs/2409.03912} {arXiv:2409.03912 [gr-qc]} \BibitemShut {NoStop}%
\bibitem [{\citenamefont {Macas}\ \emph {et~al.}(2022)\citenamefont {Macas}, \citenamefont {Pooley}, \citenamefont {Nuttall}, \citenamefont {Davis}, \citenamefont {Dyer}, \citenamefont {Lecoeuche}, \citenamefont {Lyman}, \citenamefont {McIver},\ and\ \citenamefont {Rink}}]{Macas:2022afm}%
  \BibitemOpen
  \bibfield  {author} {\bibinfo {author} {\bibfnamefont {R.}~\bibnamefont {Macas}}, \bibinfo {author} {\bibfnamefont {J.}~\bibnamefont {Pooley}}, \bibinfo {author} {\bibfnamefont {L.~K.}\ \bibnamefont {Nuttall}}, \bibinfo {author} {\bibfnamefont {D.}~\bibnamefont {Davis}}, \bibinfo {author} {\bibfnamefont {M.~J.}\ \bibnamefont {Dyer}}, \bibinfo {author} {\bibfnamefont {Y.}~\bibnamefont {Lecoeuche}}, \bibinfo {author} {\bibfnamefont {J.~D.}\ \bibnamefont {Lyman}}, \bibinfo {author} {\bibfnamefont {J.}~\bibnamefont {McIver}},\ and\ \bibinfo {author} {\bibfnamefont {K.}~\bibnamefont {Rink}},\ }\href {https://doi.org/10.1103/PhysRevD.105.103021} {\bibfield  {journal} {\bibinfo  {journal} {Phys. Rev. D}\ }\textbf {\bibinfo {volume} {105}},\ \bibinfo {pages} {103021} (\bibinfo {year} {2022})},\ \Eprint {https://arxiv.org/abs/2202.00344} {arXiv:2202.00344 [astro-ph.HE]} \BibitemShut {NoStop}%
\bibitem [{\citenamefont {Powell}(2018)}]{Powell:2018csz}%
  \BibitemOpen
  \bibfield  {author} {\bibinfo {author} {\bibfnamefont {J.}~\bibnamefont {Powell}},\ }\href {https://doi.org/10.1088/1361-6382/aacf18} {\bibfield  {journal} {\bibinfo  {journal} {Class. Quant. Grav.}\ }\textbf {\bibinfo {volume} {35}},\ \bibinfo {pages} {155017} (\bibinfo {year} {2018})},\ \Eprint {https://arxiv.org/abs/1803.11346} {arXiv:1803.11346 [astro-ph.IM]} \BibitemShut {NoStop}%
\bibitem [{\citenamefont {Kwok}\ \emph {et~al.}(2022)\citenamefont {Kwok}, \citenamefont {Lo}, \citenamefont {Weinstein},\ and\ \citenamefont {Li}}]{Kwok:2021zny}%
  \BibitemOpen
  \bibfield  {author} {\bibinfo {author} {\bibfnamefont {J.~Y.~L.}\ \bibnamefont {Kwok}}, \bibinfo {author} {\bibfnamefont {R.~K.~L.}\ \bibnamefont {Lo}}, \bibinfo {author} {\bibfnamefont {A.~J.}\ \bibnamefont {Weinstein}},\ and\ \bibinfo {author} {\bibfnamefont {T.~G.~F.}\ \bibnamefont {Li}},\ }\href {https://doi.org/10.1103/PhysRevD.105.024066} {\bibfield  {journal} {\bibinfo  {journal} {Phys. Rev. D}\ }\textbf {\bibinfo {volume} {105}},\ \bibinfo {pages} {024066} (\bibinfo {year} {2022})},\ \Eprint {https://arxiv.org/abs/2109.07642} {arXiv:2109.07642 [gr-qc]} \BibitemShut {NoStop}%
\bibitem [{\citenamefont {Mozzon}\ \emph {et~al.}(2022)\citenamefont {Mozzon}, \citenamefont {Ashton}, \citenamefont {Nuttall},\ and\ \citenamefont {Williamson}}]{Mozzon:2021wam}%
  \BibitemOpen
  \bibfield  {author} {\bibinfo {author} {\bibfnamefont {S.}~\bibnamefont {Mozzon}}, \bibinfo {author} {\bibfnamefont {G.}~\bibnamefont {Ashton}}, \bibinfo {author} {\bibfnamefont {L.~K.}\ \bibnamefont {Nuttall}},\ and\ \bibinfo {author} {\bibfnamefont {A.~R.}\ \bibnamefont {Williamson}},\ }\href {https://doi.org/10.1103/PhysRevD.106.043504} {\bibfield  {journal} {\bibinfo  {journal} {Phys. Rev. D}\ }\textbf {\bibinfo {volume} {106}},\ \bibinfo {pages} {043504} (\bibinfo {year} {2022})},\ \Eprint {https://arxiv.org/abs/2110.11731} {arXiv:2110.11731 [astro-ph.CO]} \BibitemShut {NoStop}%
\bibitem [{\citenamefont {Chatziioannou}\ \emph {et~al.}(2021)\citenamefont {Chatziioannou}, \citenamefont {Cornish}, \citenamefont {Wijngaarden},\ and\ \citenamefont {Littenberg}}]{Chatziioannou:2021ezd}%
  \BibitemOpen
  \bibfield  {author} {\bibinfo {author} {\bibfnamefont {K.}~\bibnamefont {Chatziioannou}}, \bibinfo {author} {\bibfnamefont {N.}~\bibnamefont {Cornish}}, \bibinfo {author} {\bibfnamefont {M.}~\bibnamefont {Wijngaarden}},\ and\ \bibinfo {author} {\bibfnamefont {T.~B.}\ \bibnamefont {Littenberg}},\ }\href {https://doi.org/10.1103/PhysRevD.103.044013} {\bibfield  {journal} {\bibinfo  {journal} {Phys. Rev. D}\ }\textbf {\bibinfo {volume} {103}},\ \bibinfo {pages} {044013} (\bibinfo {year} {2021})},\ \Eprint {https://arxiv.org/abs/2101.01200} {arXiv:2101.01200 [gr-qc]} \BibitemShut {NoStop}%
\bibitem [{\citenamefont {Payne}\ \emph {et~al.}(2022{\natexlab{b}})\citenamefont {Payne}, \citenamefont {Hourihane}, \citenamefont {Golomb}, \citenamefont {Udall}, \citenamefont {Davis},\ and\ \citenamefont {Chatziioannou}}]{Payne:2022spz}%
  \BibitemOpen
  \bibfield  {author} {\bibinfo {author} {\bibfnamefont {E.}~\bibnamefont {Payne}}, \bibinfo {author} {\bibfnamefont {S.}~\bibnamefont {Hourihane}}, \bibinfo {author} {\bibfnamefont {J.}~\bibnamefont {Golomb}}, \bibinfo {author} {\bibfnamefont {R.}~\bibnamefont {Udall}}, \bibinfo {author} {\bibfnamefont {D.}~\bibnamefont {Davis}},\ and\ \bibinfo {author} {\bibfnamefont {K.}~\bibnamefont {Chatziioannou}},\ }\href {https://doi.org/10.1103/PhysRevD.106.104017} {\bibfield  {journal} {\bibinfo  {journal} {Phys. Rev. D}\ }\textbf {\bibinfo {volume} {106}},\ \bibinfo {pages} {104017} (\bibinfo {year} {2022}{\natexlab{b}})},\ \Eprint {https://arxiv.org/abs/2206.11932} {arXiv:2206.11932 [gr-qc]} \BibitemShut {NoStop}%
\bibitem [{\citenamefont {Abbott}\ \emph {et~al.}(2017)\citenamefont {Abbott} \emph {et~al.}}]{TheLIGOScientific:2017qsa}%
  \BibitemOpen
  \bibfield  {author} {\bibinfo {author} {\bibfnamefont {B.~P.}\ \bibnamefont {Abbott}} \emph {et~al.} (\bibinfo {collaboration} {LIGO Scientific Collaboration, Virgo Collaboration}),\ }\href {https://doi.org/10.1103/PhysRevLett.119.161101} {\bibfield  {journal} {\bibinfo  {journal} {Phys. Rev. Lett.}\ }\textbf {\bibinfo {volume} {119}},\ \bibinfo {pages} {161101} (\bibinfo {year} {2017})},\ \Eprint {https://arxiv.org/abs/1710.05832} {arXiv:1710.05832 [gr-qc]} \BibitemShut {NoStop}%
\bibitem [{\citenamefont {Abbott}\ \emph {et~al.}(2021{\natexlab{c}})\citenamefont {Abbott} \emph {et~al.}}]{GWTC2.1}%
  \BibitemOpen
  \bibfield  {author} {\bibinfo {author} {\bibfnamefont {R.}~\bibnamefont {Abbott}} \emph {et~al.} (\bibinfo {collaboration} {LIGO Scientific, VIRGO}),\ }\href@noop {} {\bibinfo {title} {{GWTC-2.1: Deep Extended Catalog of Compact Binary Coalescences Observed by LIGO and Virgo During the First Half of the Third Observing Run}}} (\bibinfo {year} {2021}{\natexlab{c}}),\ \Eprint {https://arxiv.org/abs/2108.01045} {arXiv:2108.01045 [gr-qc]} \BibitemShut {NoStop}%
\bibitem [{\citenamefont {Davis}\ \emph {et~al.}(2022)\citenamefont {Davis}, \citenamefont {Littenberg}, \citenamefont {Romero-Shaw}, \citenamefont {Millhouse}, \citenamefont {McIver}, \citenamefont {Di~Renzo},\ and\ \citenamefont {Ashton}}]{Davis:2022ird}%
  \BibitemOpen
  \bibfield  {author} {\bibinfo {author} {\bibfnamefont {D.}~\bibnamefont {Davis}}, \bibinfo {author} {\bibfnamefont {T.~B.}\ \bibnamefont {Littenberg}}, \bibinfo {author} {\bibfnamefont {I.~M.}\ \bibnamefont {Romero-Shaw}}, \bibinfo {author} {\bibfnamefont {M.}~\bibnamefont {Millhouse}}, \bibinfo {author} {\bibfnamefont {J.}~\bibnamefont {McIver}}, \bibinfo {author} {\bibfnamefont {F.}~\bibnamefont {Di~Renzo}},\ and\ \bibinfo {author} {\bibfnamefont {G.}~\bibnamefont {Ashton}},\ }\href {https://doi.org/10.1088/1361-6382/aca238} {\bibfield  {journal} {\bibinfo  {journal} {Class. Quant. Grav.}\ }\textbf {\bibinfo {volume} {39}},\ \bibinfo {pages} {245013} (\bibinfo {year} {2022})},\ \Eprint {https://arxiv.org/abs/2207.03429} {arXiv:2207.03429 [astro-ph.IM]} \BibitemShut {NoStop}%
\bibitem [{\citenamefont {Vazsonyi}\ and\ \citenamefont {Davis}(2023)}]{Vazsonyi:2022jul}%
  \BibitemOpen
  \bibfield  {author} {\bibinfo {author} {\bibfnamefont {L.}~\bibnamefont {Vazsonyi}}\ and\ \bibinfo {author} {\bibfnamefont {D.}~\bibnamefont {Davis}},\ }\href {https://doi.org/10.1088/1361-6382/acafd2} {\bibfield  {journal} {\bibinfo  {journal} {Class. Quant. Grav.}\ }\textbf {\bibinfo {volume} {40}},\ \bibinfo {pages} {035008} (\bibinfo {year} {2023})},\ \Eprint {https://arxiv.org/abs/2208.12338} {arXiv:2208.12338 [astro-ph.IM]} \BibitemShut {NoStop}%
\bibitem [{\citenamefont {Miller}\ \emph {et~al.}(2024)\citenamefont {Miller}, \citenamefont {Isi}, \citenamefont {Chatziioannou}, \citenamefont {Varma},\ and\ \citenamefont {Mandel}}]{Miller:2023ncs}%
  \BibitemOpen
  \bibfield  {author} {\bibinfo {author} {\bibfnamefont {S.~J.}\ \bibnamefont {Miller}}, \bibinfo {author} {\bibfnamefont {M.}~\bibnamefont {Isi}}, \bibinfo {author} {\bibfnamefont {K.}~\bibnamefont {Chatziioannou}}, \bibinfo {author} {\bibfnamefont {V.}~\bibnamefont {Varma}},\ and\ \bibinfo {author} {\bibfnamefont {I.}~\bibnamefont {Mandel}},\ }\href {https://doi.org/10.1103/PhysRevD.109.024024} {\bibfield  {journal} {\bibinfo  {journal} {Phys. Rev. D}\ }\textbf {\bibinfo {volume} {109}},\ \bibinfo {pages} {024024} (\bibinfo {year} {2024})},\ \Eprint {https://arxiv.org/abs/2310.01544} {arXiv:2310.01544 [astro-ph.HE]} \BibitemShut {NoStop}%
\bibitem [{\citenamefont {Abbott}\ \emph {et~al.}(2021{\natexlab{d}})\citenamefont {Abbott} \emph {et~al.}}]{LIGOScientific:2020ibl}%
  \BibitemOpen
  \bibfield  {author} {\bibinfo {author} {\bibfnamefont {R.}~\bibnamefont {Abbott}} \emph {et~al.} (\bibinfo {collaboration} {LIGO Scientific, Virgo}),\ }\href {https://doi.org/10.1103/PhysRevX.11.021053} {\bibfield  {journal} {\bibinfo  {journal} {Phys. Rev. X}\ }\textbf {\bibinfo {volume} {11}},\ \bibinfo {pages} {021053} (\bibinfo {year} {2021}{\natexlab{d}})},\ \Eprint {https://arxiv.org/abs/2010.14527} {arXiv:2010.14527 [gr-qc]} \BibitemShut {NoStop}%
\bibitem [{\citenamefont {Davis}\ \emph {et~al.}(2019)\citenamefont {Davis}, \citenamefont {Massinger}, \citenamefont {Lundgren}, \citenamefont {Driggers}, \citenamefont {Urban},\ and\ \citenamefont {Nuttall}}]{Davis:2018yrz}%
  \BibitemOpen
  \bibfield  {author} {\bibinfo {author} {\bibfnamefont {D.}~\bibnamefont {Davis}}, \bibinfo {author} {\bibfnamefont {T.~J.}\ \bibnamefont {Massinger}}, \bibinfo {author} {\bibfnamefont {A.~P.}\ \bibnamefont {Lundgren}}, \bibinfo {author} {\bibfnamefont {J.~C.}\ \bibnamefont {Driggers}}, \bibinfo {author} {\bibfnamefont {A.~L.}\ \bibnamefont {Urban}},\ and\ \bibinfo {author} {\bibfnamefont {L.~K.}\ \bibnamefont {Nuttall}},\ }\href {https://doi.org/10.1088/1361-6382/ab01c5} {\bibfield  {journal} {\bibinfo  {journal} {Class. Quant. Grav.}\ }\textbf {\bibinfo {volume} {36}},\ \bibinfo {pages} {055011} (\bibinfo {year} {2019})},\ \Eprint {https://arxiv.org/abs/1809.05348} {arXiv:1809.05348 [astro-ph.IM]} \BibitemShut {NoStop}%
\bibitem [{\citenamefont {Chatziioannou}\ \emph {et~al.}(2019{\natexlab{a}})\citenamefont {Chatziioannou}, \citenamefont {Haster}, \citenamefont {Littenberg}, \citenamefont {Farr}, \citenamefont {Ghonge}, \citenamefont {Millhouse}, \citenamefont {Clark},\ and\ \citenamefont {Cornish}}]{Chatziioannou:2019zvs}%
  \BibitemOpen
  \bibfield  {author} {\bibinfo {author} {\bibfnamefont {K.}~\bibnamefont {Chatziioannou}}, \bibinfo {author} {\bibfnamefont {C.-J.}\ \bibnamefont {Haster}}, \bibinfo {author} {\bibfnamefont {T.~B.}\ \bibnamefont {Littenberg}}, \bibinfo {author} {\bibfnamefont {W.~M.}\ \bibnamefont {Farr}}, \bibinfo {author} {\bibfnamefont {S.}~\bibnamefont {Ghonge}}, \bibinfo {author} {\bibfnamefont {M.}~\bibnamefont {Millhouse}}, \bibinfo {author} {\bibfnamefont {J.~A.}\ \bibnamefont {Clark}},\ and\ \bibinfo {author} {\bibfnamefont {N.}~\bibnamefont {Cornish}},\ }\href {https://doi.org/10.1103/PhysRevD.100.104004} {\bibfield  {journal} {\bibinfo  {journal} {Phys. Rev. D}\ }\textbf {\bibinfo {volume} {100}},\ \bibinfo {pages} {104004} (\bibinfo {year} {2019}{\natexlab{a}})},\ \Eprint {https://arxiv.org/abs/1907.06540} {arXiv:1907.06540 [gr-qc]} \BibitemShut {NoStop}%
\bibitem [{\citenamefont {Zweizig}\ and\ \citenamefont {Riles}(2021)}]{Zweizig:gating}%
  \BibitemOpen
  \bibfield  {author} {\bibinfo {author} {\bibfnamefont {J.}~\bibnamefont {Zweizig}}\ and\ \bibinfo {author} {\bibfnamefont {K.}~\bibnamefont {Riles}},\ }\href@noop {} {\emph {\bibinfo {title} {{Information on self-gating of $h(t)$ used in O3 continuous-wave and stochastic searches}}}},\ \bibinfo {type} {Tech. Rep.}\ (\bibinfo {address} {https://dcc.ligo.org/LIGO-T2000384},\ \bibinfo {year} {2021})\BibitemShut {NoStop}%
\bibitem [{\citenamefont {Zackay}\ \emph {et~al.}(2021)\citenamefont {Zackay}, \citenamefont {Venumadhav}, \citenamefont {Roulet}, \citenamefont {Dai},\ and\ \citenamefont {Zaldarriaga}}]{Zackay:2019kkv}%
  \BibitemOpen
  \bibfield  {author} {\bibinfo {author} {\bibfnamefont {B.}~\bibnamefont {Zackay}}, \bibinfo {author} {\bibfnamefont {T.}~\bibnamefont {Venumadhav}}, \bibinfo {author} {\bibfnamefont {J.}~\bibnamefont {Roulet}}, \bibinfo {author} {\bibfnamefont {L.}~\bibnamefont {Dai}},\ and\ \bibinfo {author} {\bibfnamefont {M.}~\bibnamefont {Zaldarriaga}},\ }\href {https://doi.org/10.1103/PhysRevD.104.063034} {\bibfield  {journal} {\bibinfo  {journal} {Phys. Rev. D}\ }\textbf {\bibinfo {volume} {104}},\ \bibinfo {pages} {063034} (\bibinfo {year} {2021})},\ \Eprint {https://arxiv.org/abs/1908.05644} {arXiv:1908.05644 [astro-ph.IM]} \BibitemShut {NoStop}%
\bibitem [{\citenamefont {Soni}\ \emph {et~al.}(2024)\citenamefont {Soni}, \citenamefont {Glanzer}, \citenamefont {Effler}, \citenamefont {Frolov}, \citenamefont {Gonz\'alez}, \citenamefont {Pele},\ and\ \citenamefont {Schofield}}]{Soni:2023kqq}%
  \BibitemOpen
  \bibfield  {author} {\bibinfo {author} {\bibfnamefont {S.}~\bibnamefont {Soni}}, \bibinfo {author} {\bibfnamefont {J.}~\bibnamefont {Glanzer}}, \bibinfo {author} {\bibfnamefont {A.}~\bibnamefont {Effler}}, \bibinfo {author} {\bibfnamefont {V.}~\bibnamefont {Frolov}}, \bibinfo {author} {\bibfnamefont {G.}~\bibnamefont {Gonz\'alez}}, \bibinfo {author} {\bibfnamefont {A.}~\bibnamefont {Pele}},\ and\ \bibinfo {author} {\bibfnamefont {R.}~\bibnamefont {Schofield}},\ }\href {https://doi.org/10.1088/1361-6382/ad494a} {\bibfield  {journal} {\bibinfo  {journal} {Class. Quant. Grav.}\ }\textbf {\bibinfo {volume} {41}},\ \bibinfo {pages} {135015} (\bibinfo {year} {2024})},\ \Eprint {https://arxiv.org/abs/2311.05730} {arXiv:2311.05730 [astro-ph.IM]} \BibitemShut {NoStop}%
\bibitem [{\citenamefont {Soni}\ \emph {et~al.}(2020)\citenamefont {Soni} \emph {et~al.}}]{LIGO:2020zwl}%
  \BibitemOpen
  \bibfield  {author} {\bibinfo {author} {\bibfnamefont {S.}~\bibnamefont {Soni}} \emph {et~al.} (\bibinfo {collaboration} {LIGO}),\ }\href {https://doi.org/10.1088/1361-6382/abc906} {\bibfield  {journal} {\bibinfo  {journal} {Class. Quant. Grav.}\ }\textbf {\bibinfo {volume} {38}},\ \bibinfo {pages} {025016} (\bibinfo {year} {2020})},\ \Eprint {https://arxiv.org/abs/2007.14876} {arXiv:2007.14876 [astro-ph.IM]} \BibitemShut {NoStop}%
\bibitem [{\citenamefont {Udall}\ and\ \citenamefont {Davis}(2023)}]{Udall:2022vkv}%
  \BibitemOpen
  \bibfield  {author} {\bibinfo {author} {\bibfnamefont {R.}~\bibnamefont {Udall}}\ and\ \bibinfo {author} {\bibfnamefont {D.}~\bibnamefont {Davis}},\ }\href {https://doi.org/10.1063/5.0136896} {\bibfield  {journal} {\bibinfo  {journal} {Appl. Phys. Lett.}\ }\textbf {\bibinfo {volume} {122}},\ \bibinfo {pages} {094103} (\bibinfo {year} {2023})},\ \Eprint {https://arxiv.org/abs/2211.15867} {arXiv:2211.15867 [astro-ph.IM]} \BibitemShut {NoStop}%
\bibitem [{\citenamefont {Cornish}\ \emph {et~al.}(2021)\citenamefont {Cornish}, \citenamefont {Littenberg}, \citenamefont {B\'ecsy}, \citenamefont {Chatziioannou}, \citenamefont {Clark}, \citenamefont {Ghonge},\ and\ \citenamefont {Millhouse}}]{Cornish:2020dwh}%
  \BibitemOpen
  \bibfield  {author} {\bibinfo {author} {\bibfnamefont {N.~J.}\ \bibnamefont {Cornish}}, \bibinfo {author} {\bibfnamefont {T.~B.}\ \bibnamefont {Littenberg}}, \bibinfo {author} {\bibfnamefont {B.}~\bibnamefont {B\'ecsy}}, \bibinfo {author} {\bibfnamefont {K.}~\bibnamefont {Chatziioannou}}, \bibinfo {author} {\bibfnamefont {J.~A.}\ \bibnamefont {Clark}}, \bibinfo {author} {\bibfnamefont {S.}~\bibnamefont {Ghonge}},\ and\ \bibinfo {author} {\bibfnamefont {M.}~\bibnamefont {Millhouse}},\ }\href {https://doi.org/10.1103/PhysRevD.103.044006} {\bibfield  {journal} {\bibinfo  {journal} {Phys. Rev. D}\ }\textbf {\bibinfo {volume} {103}},\ \bibinfo {pages} {044006} (\bibinfo {year} {2021})},\ \Eprint {https://arxiv.org/abs/2011.09494} {arXiv:2011.09494 [gr-qc]} \BibitemShut {NoStop}%
\bibitem [{\citenamefont {Cornish}\ and\ \citenamefont {Littenberg}(2015)}]{Cornish:2014kda}%
  \BibitemOpen
  \bibfield  {author} {\bibinfo {author} {\bibfnamefont {N.~J.}\ \bibnamefont {Cornish}}\ and\ \bibinfo {author} {\bibfnamefont {T.~B.}\ \bibnamefont {Littenberg}},\ }\href {https://doi.org/10.1088/0264-9381/32/13/135012} {\bibfield  {journal} {\bibinfo  {journal} {Class. Quant. Grav.}\ }\textbf {\bibinfo {volume} {32}},\ \bibinfo {pages} {135012} (\bibinfo {year} {2015})},\ \Eprint {https://arxiv.org/abs/1410.3835} {arXiv:1410.3835 [gr-qc]} \BibitemShut {NoStop}%
\bibitem [{\citenamefont {Romano}\ and\ \citenamefont {Cornish}(2017)}]{Romano:2016dpx}%
  \BibitemOpen
  \bibfield  {author} {\bibinfo {author} {\bibfnamefont {J.~D.}\ \bibnamefont {Romano}}\ and\ \bibinfo {author} {\bibfnamefont {N.~J.}\ \bibnamefont {Cornish}},\ }\href {https://doi.org/10.1007/s41114-017-0004-1} {\bibfield  {journal} {\bibinfo  {journal} {Living Rev. Rel.}\ }\textbf {\bibinfo {volume} {20}},\ \bibinfo {pages} {2} (\bibinfo {year} {2017})},\ \Eprint {https://arxiv.org/abs/1608.06889} {arXiv:1608.06889 [gr-qc]} \BibitemShut {NoStop}%
\bibitem [{\citenamefont {Deshong}\ \emph {et~al.}(2025)\citenamefont {Deshong}, \citenamefont {Hourihane}, \citenamefont {Chatziioannou}, \citenamefont {Johnson}, \citenamefont {Holst}, \citenamefont {Millhouse}, \citenamefont {Littenberg},\ and\ \citenamefont {Cornish}}]{bayeswave_cpp}%
  \BibitemOpen
  \bibfield  {author} {\bibinfo {author} {\bibfnamefont {H.}~\bibnamefont {Deshong}}, \bibinfo {author} {\bibfnamefont {S.}~\bibnamefont {Hourihane}}, \bibinfo {author} {\bibfnamefont {K.}~\bibnamefont {Chatziioannou}}, \bibinfo {author} {\bibfnamefont {A.}~\bibnamefont {Johnson}}, \bibinfo {author} {\bibfnamefont {M.}~\bibnamefont {Holst}}, \bibinfo {author} {\bibfnamefont {M.}~\bibnamefont {Millhouse}}, \bibinfo {author} {\bibfnamefont {T.}~\bibnamefont {Littenberg}},\ and\ \bibinfo {author} {\bibfnamefont {N.}~\bibnamefont {Cornish}},\ }\href {https://git.ligo.org/bayeswave/bayeswave-cpp} {\bibinfo {title} {Bayeswave c++ repository}} (\bibinfo {year} {2025}),\ \bibinfo {note} {gitLab repository, accessed May 21, 2025}\BibitemShut {NoStop}%
\bibitem [{\citenamefont {{LIGO Scientific Collaboration and Virgo Collaboration}}(2018)}]{bayeswave}%
  \BibitemOpen
  \bibfield  {author} {\bibinfo {author} {\bibnamefont {{LIGO Scientific Collaboration and Virgo Collaboration}}},\ }\href {https://git.ligo.org/lscsoft/bayeswave} {\bibinfo {title} {{BayesWave}, https://git.ligo.org/lscsoft/bayeswave}} (\bibinfo {year} {2018})\BibitemShut {NoStop}%
\bibitem [{\citenamefont {Pratten}\ \emph {et~al.}(2021)\citenamefont {Pratten} \emph {et~al.}}]{Pratten:2020ceb}%
  \BibitemOpen
  \bibfield  {author} {\bibinfo {author} {\bibfnamefont {G.}~\bibnamefont {Pratten}} \emph {et~al.},\ }\href {https://doi.org/10.1103/PhysRevD.103.104056} {\bibfield  {journal} {\bibinfo  {journal} {Phys. Rev. D}\ }\textbf {\bibinfo {volume} {103}},\ \bibinfo {pages} {104056} (\bibinfo {year} {2021})},\ \Eprint {https://arxiv.org/abs/2004.06503} {arXiv:2004.06503 [gr-qc]} \BibitemShut {NoStop}%
\bibitem [{\citenamefont {Schmidt}\ \emph {et~al.}(2015)\citenamefont {Schmidt}, \citenamefont {Ohme},\ and\ \citenamefont {Hannam}}]{chi_p}%
  \BibitemOpen
  \bibfield  {author} {\bibinfo {author} {\bibfnamefont {P.}~\bibnamefont {Schmidt}}, \bibinfo {author} {\bibfnamefont {F.}~\bibnamefont {Ohme}},\ and\ \bibinfo {author} {\bibfnamefont {M.}~\bibnamefont {Hannam}},\ }\bibfield  {journal} {\bibinfo  {journal} {Physical Review D}\ }\textbf {\bibinfo {volume} {91}},\ \href {https://doi.org/10.1103/physrevd.91.024043} {10.1103/physrevd.91.024043} (\bibinfo {year} {2015})\BibitemShut {NoStop}%
\bibitem [{\citenamefont {Hourihane}\ \emph {et~al.}(2023)\citenamefont {Hourihane}, \citenamefont {Meyers}, \citenamefont {Johnson}, \citenamefont {Chatziioannou},\ and\ \citenamefont {Vallisneri}}]{Hourihane:2022ner_PTA}%
  \BibitemOpen
  \bibfield  {author} {\bibinfo {author} {\bibfnamefont {S.}~\bibnamefont {Hourihane}}, \bibinfo {author} {\bibfnamefont {P.}~\bibnamefont {Meyers}}, \bibinfo {author} {\bibfnamefont {A.}~\bibnamefont {Johnson}}, \bibinfo {author} {\bibfnamefont {K.}~\bibnamefont {Chatziioannou}},\ and\ \bibinfo {author} {\bibfnamefont {M.}~\bibnamefont {Vallisneri}},\ }\href {https://doi.org/10.1103/PhysRevD.107.084045} {\bibfield  {journal} {\bibinfo  {journal} {Phys. Rev. D}\ }\textbf {\bibinfo {volume} {107}},\ \bibinfo {pages} {084045} (\bibinfo {year} {2023})},\ \Eprint {https://arxiv.org/abs/2212.06276} {arXiv:2212.06276 [gr-qc]} \BibitemShut {NoStop}%
\bibitem [{\citenamefont {Payne}\ \emph {et~al.}(2019)\citenamefont {Payne}, \citenamefont {Talbot},\ and\ \citenamefont {Thrane}}]{Payne:2019wmy}%
  \BibitemOpen
  \bibfield  {author} {\bibinfo {author} {\bibfnamefont {E.}~\bibnamefont {Payne}}, \bibinfo {author} {\bibfnamefont {C.}~\bibnamefont {Talbot}},\ and\ \bibinfo {author} {\bibfnamefont {E.}~\bibnamefont {Thrane}},\ }\href {https://doi.org/10.1103/PhysRevD.100.123017} {\bibfield  {journal} {\bibinfo  {journal} {Phys. Rev. D}\ }\textbf {\bibinfo {volume} {100}},\ \bibinfo {pages} {123017} (\bibinfo {year} {2019})},\ \Eprint {https://arxiv.org/abs/1905.05477} {arXiv:1905.05477 [astro-ph.IM]} \BibitemShut {NoStop}%
\bibitem [{\citenamefont {Romero-Shaw}\ \emph {et~al.}(2020{\natexlab{a}})\citenamefont {Romero-Shaw}, \citenamefont {Lasky}, \citenamefont {Thrane},\ and\ \citenamefont {Bustillo}}]{Romero-Shaw:2020thy}%
  \BibitemOpen
  \bibfield  {author} {\bibinfo {author} {\bibfnamefont {I.~M.}\ \bibnamefont {Romero-Shaw}}, \bibinfo {author} {\bibfnamefont {P.~D.}\ \bibnamefont {Lasky}}, \bibinfo {author} {\bibfnamefont {E.}~\bibnamefont {Thrane}},\ and\ \bibinfo {author} {\bibfnamefont {J.~C.}\ \bibnamefont {Bustillo}},\ }\href {https://doi.org/10.3847/2041-8213/abbe26} {\bibfield  {journal} {\bibinfo  {journal} {Astrophys. J. Lett.}\ }\textbf {\bibinfo {volume} {903}},\ \bibinfo {pages} {L5} (\bibinfo {year} {2020}{\natexlab{a}})},\ \Eprint {https://arxiv.org/abs/2009.04771} {arXiv:2009.04771 [astro-ph.HE]} \BibitemShut {NoStop}%
\bibitem [{\citenamefont {Romero-Shaw}\ \emph {et~al.}(2022)\citenamefont {Romero-Shaw}, \citenamefont {Lasky},\ and\ \citenamefont {Thrane}}]{Romero-Shaw:2022xko}%
  \BibitemOpen
  \bibfield  {author} {\bibinfo {author} {\bibfnamefont {I.~M.}\ \bibnamefont {Romero-Shaw}}, \bibinfo {author} {\bibfnamefont {P.~D.}\ \bibnamefont {Lasky}},\ and\ \bibinfo {author} {\bibfnamefont {E.}~\bibnamefont {Thrane}},\ }\href {https://doi.org/10.3847/1538-4357/ac9798} {\bibfield  {journal} {\bibinfo  {journal} {Astrophys. J.}\ }\textbf {\bibinfo {volume} {940}},\ \bibinfo {pages} {171} (\bibinfo {year} {2022})},\ \Eprint {https://arxiv.org/abs/2206.14695} {arXiv:2206.14695 [astro-ph.HE]} \BibitemShut {NoStop}%
\bibitem [{\citenamefont {Romero-Shaw}\ \emph {et~al.}(2020{\natexlab{b}})\citenamefont {Romero-Shaw}, \citenamefont {Farrow}, \citenamefont {Stevenson}, \citenamefont {Thrane},\ and\ \citenamefont {Zhu}}]{Romero-Shaw:2020aaj}%
  \BibitemOpen
  \bibfield  {author} {\bibinfo {author} {\bibfnamefont {I.~M.}\ \bibnamefont {Romero-Shaw}}, \bibinfo {author} {\bibfnamefont {N.}~\bibnamefont {Farrow}}, \bibinfo {author} {\bibfnamefont {S.}~\bibnamefont {Stevenson}}, \bibinfo {author} {\bibfnamefont {E.}~\bibnamefont {Thrane}},\ and\ \bibinfo {author} {\bibfnamefont {X.-J.}\ \bibnamefont {Zhu}},\ }\href {https://doi.org/10.1093/mnrasl/slaa084} {\bibfield  {journal} {\bibinfo  {journal} {Mon. Not. Roy. Astron. Soc.}\ }\textbf {\bibinfo {volume} {496}},\ \bibinfo {pages} {L64} (\bibinfo {year} {2020}{\natexlab{b}})},\ \Eprint {https://arxiv.org/abs/2001.06492} {arXiv:2001.06492 [astro-ph.HE]} \BibitemShut {NoStop}%
\bibitem [{\citenamefont {Romero-Shaw}\ \emph {et~al.}(2021)\citenamefont {Romero-Shaw}, \citenamefont {Lasky},\ and\ \citenamefont {Thrane}}]{Romero-Shaw:2021ual}%
  \BibitemOpen
  \bibfield  {author} {\bibinfo {author} {\bibfnamefont {I.~M.}\ \bibnamefont {Romero-Shaw}}, \bibinfo {author} {\bibfnamefont {P.~D.}\ \bibnamefont {Lasky}},\ and\ \bibinfo {author} {\bibfnamefont {E.}~\bibnamefont {Thrane}},\ }\href {https://doi.org/10.3847/2041-8213/ac3138} {\bibfield  {journal} {\bibinfo  {journal} {Astrophys. J. Lett.}\ }\textbf {\bibinfo {volume} {921}},\ \bibinfo {pages} {L31} (\bibinfo {year} {2021})},\ \Eprint {https://arxiv.org/abs/2108.01284} {arXiv:2108.01284 [astro-ph.HE]} \BibitemShut {NoStop}%
\bibitem [{\citenamefont {Dax}\ \emph {et~al.}(2025)\citenamefont {Dax}, \citenamefont {Green}, \citenamefont {Gair}, \citenamefont {Gupte}, \citenamefont {P\"urrer}, \citenamefont {Raymond}, \citenamefont {Wildberger}, \citenamefont {Macke}, \citenamefont {Buonanno},\ and\ \citenamefont {Sch\"olkopf}}]{Dax:2024mcn}%
  \BibitemOpen
  \bibfield  {author} {\bibinfo {author} {\bibfnamefont {M.}~\bibnamefont {Dax}}, \bibinfo {author} {\bibfnamefont {S.~R.}\ \bibnamefont {Green}}, \bibinfo {author} {\bibfnamefont {J.}~\bibnamefont {Gair}}, \bibinfo {author} {\bibfnamefont {N.}~\bibnamefont {Gupte}}, \bibinfo {author} {\bibfnamefont {M.}~\bibnamefont {P\"urrer}}, \bibinfo {author} {\bibfnamefont {V.}~\bibnamefont {Raymond}}, \bibinfo {author} {\bibfnamefont {J.}~\bibnamefont {Wildberger}}, \bibinfo {author} {\bibfnamefont {J.~H.}\ \bibnamefont {Macke}}, \bibinfo {author} {\bibfnamefont {A.}~\bibnamefont {Buonanno}},\ and\ \bibinfo {author} {\bibfnamefont {B.}~\bibnamefont {Sch\"olkopf}},\ }\href {https://doi.org/10.1038/s41586-025-08593-z} {\bibfield  {journal} {\bibinfo  {journal} {Nature}\ }\textbf {\bibinfo {volume} {639}},\ \bibinfo {pages} {49} (\bibinfo {year} {2025})},\ \Eprint {https://arxiv.org/abs/2407.09602} {arXiv:2407.09602 [gr-qc]} \BibitemShut {NoStop}%
\bibitem [{\citenamefont {Lin}(1991)}]{Lin:1991}%
  \BibitemOpen
  \bibfield  {author} {\bibinfo {author} {\bibfnamefont {J.}~\bibnamefont {Lin}},\ }\href {https://doi.org/10.1109/18.61115} {\bibfield  {journal} {\bibinfo  {journal} {IEEE Transactions on Information Theory}\ }\textbf {\bibinfo {volume} {37}},\ \bibinfo {pages} {145} (\bibinfo {year} {1991})}\BibitemShut {NoStop}%
\bibitem [{\citenamefont {Romero-Shaw}\ \emph {et~al.}(2020{\natexlab{c}})\citenamefont {Romero-Shaw} \emph {et~al.}}]{Romero-Shaw:2020owr}%
  \BibitemOpen
  \bibfield  {author} {\bibinfo {author} {\bibfnamefont {I.~M.}\ \bibnamefont {Romero-Shaw}} \emph {et~al.},\ }\href {https://doi.org/10.1093/mnras/staa2850} {\bibfield  {journal} {\bibinfo  {journal} {Mon. Not. Roy. Astron. Soc.}\ }\textbf {\bibinfo {volume} {499}},\ \bibinfo {pages} {3295} (\bibinfo {year} {2020}{\natexlab{c}})},\ \Eprint {https://arxiv.org/abs/2006.00714} {arXiv:2006.00714 [astro-ph.IM]} \BibitemShut {NoStop}%
\bibitem [{\citenamefont {Gibbons}\ and\ \citenamefont {Chakraborti}(2003)}]{Gibbons2003}%
  \BibitemOpen
  \bibfield  {author} {\bibinfo {author} {\bibfnamefont {J.~D.}\ \bibnamefont {Gibbons}}\ and\ \bibinfo {author} {\bibfnamefont {S.}~\bibnamefont {Chakraborti}},\ }\href@noop {} {\emph {\bibinfo {title} {Nonparametric Statistical Inference, Fourth Edition}}}\ (\bibinfo  {publisher} {Marcel Dekker, Inc},\ \bibinfo {address} {270 Madison Avenue, New York, NY 10016},\ \bibinfo {year} {2003})\ p.\ \bibinfo {pages} {145}\BibitemShut {NoStop}%
\bibitem [{\citenamefont {Cook}\ \emph {et~al.}(2006)\citenamefont {Cook}, \citenamefont {Gelman},\ and\ \citenamefont {Rubin}}]{Cook_PP_test}%
  \BibitemOpen
  \bibfield  {author} {\bibinfo {author} {\bibfnamefont {S.~R.}\ \bibnamefont {Cook}}, \bibinfo {author} {\bibfnamefont {A.}~\bibnamefont {Gelman}},\ and\ \bibinfo {author} {\bibfnamefont {D.~B.}\ \bibnamefont {Rubin}},\ }\href {http://www.jstor.org/stable/27594203} {\bibfield  {journal} {\bibinfo  {journal} {Journal of Computational and Graphical Statistics}\ }\textbf {\bibinfo {volume} {15}},\ \bibinfo {pages} {675} (\bibinfo {year} {2006})}\BibitemShut {NoStop}%
\bibitem [{\citenamefont {Littenberg}\ and\ \citenamefont {Cornish}(2015)}]{Littenberg:2014oda}%
  \BibitemOpen
  \bibfield  {author} {\bibinfo {author} {\bibfnamefont {T.~B.}\ \bibnamefont {Littenberg}}\ and\ \bibinfo {author} {\bibfnamefont {N.~J.}\ \bibnamefont {Cornish}},\ }\href {https://doi.org/10.1103/PhysRevD.91.084034} {\bibfield  {journal} {\bibinfo  {journal} {Phys. Rev.}\ }\textbf {\bibinfo {volume} {D91}},\ \bibinfo {pages} {084034} (\bibinfo {year} {2015})},\ \Eprint {https://arxiv.org/abs/1410.3852} {arXiv:1410.3852 [gr-qc]} \BibitemShut {NoStop}%
\bibitem [{\citenamefont {Chatziioannou}\ \emph {et~al.}(2019{\natexlab{b}})\citenamefont {Chatziioannou}, \citenamefont {Haster}, \citenamefont {Littenberg}, \citenamefont {Farr}, \citenamefont {Ghonge}, \citenamefont {Millhouse}, \citenamefont {Clark},\ and\ \citenamefont {Cornish}}]{Chatziioannou:2019}%
  \BibitemOpen
  \bibfield  {author} {\bibinfo {author} {\bibfnamefont {K.}~\bibnamefont {Chatziioannou}}, \bibinfo {author} {\bibfnamefont {C.-J.}\ \bibnamefont {Haster}}, \bibinfo {author} {\bibfnamefont {T.~B.}\ \bibnamefont {Littenberg}}, \bibinfo {author} {\bibfnamefont {W.~M.}\ \bibnamefont {Farr}}, \bibinfo {author} {\bibfnamefont {S.}~\bibnamefont {Ghonge}}, \bibinfo {author} {\bibfnamefont {M.}~\bibnamefont {Millhouse}}, \bibinfo {author} {\bibfnamefont {J.~A.}\ \bibnamefont {Clark}},\ and\ \bibinfo {author} {\bibfnamefont {N.}~\bibnamefont {Cornish}},\ }\href {https://doi.org/10.1103/PhysRevD.100.104004} {\bibfield  {journal} {\bibinfo  {journal} {Phys. Rev. D}\ }\textbf {\bibinfo {volume} {100}},\ \bibinfo {pages} {104004} (\bibinfo {year} {2019}{\natexlab{b}})}\BibitemShut {NoStop}%
\bibitem [{\citenamefont {Plunkett}\ \emph {et~al.}(2022)\citenamefont {Plunkett}, \citenamefont {Hourihane},\ and\ \citenamefont {Chatziioannou}}]{Plunkett:2022zmx}%
  \BibitemOpen
  \bibfield  {author} {\bibinfo {author} {\bibfnamefont {C.}~\bibnamefont {Plunkett}}, \bibinfo {author} {\bibfnamefont {S.}~\bibnamefont {Hourihane}},\ and\ \bibinfo {author} {\bibfnamefont {K.}~\bibnamefont {Chatziioannou}},\ }\href {https://doi.org/10.1103/PhysRevD.106.104021} {\bibfield  {journal} {\bibinfo  {journal} {Phys. Rev. D}\ }\textbf {\bibinfo {volume} {106}},\ \bibinfo {pages} {104021} (\bibinfo {year} {2022})},\ \Eprint {https://arxiv.org/abs/2208.02291} {arXiv:2208.02291 [gr-qc]} \BibitemShut {NoStop}%
\bibitem [{\citenamefont {Macleod}\ \emph {et~al.}(2020)\citenamefont {Macleod}, \citenamefont {Urban}, \citenamefont {Coughlin}, \citenamefont {Massinger}, \citenamefont {Pitkin}, \citenamefont {paulaltin}, \citenamefont {Areeda}, \citenamefont {Quintero}, \citenamefont {Badger}, \citenamefont {Singer},\ and\ \citenamefont {Leinweber}}]{duncan_macleod_2020_3598469}%
  \BibitemOpen
  \bibfield  {author} {\bibinfo {author} {\bibfnamefont {D.}~\bibnamefont {Macleod}}, \bibinfo {author} {\bibfnamefont {A.~L.}\ \bibnamefont {Urban}}, \bibinfo {author} {\bibfnamefont {S.}~\bibnamefont {Coughlin}}, \bibinfo {author} {\bibfnamefont {T.}~\bibnamefont {Massinger}}, \bibinfo {author} {\bibfnamefont {M.}~\bibnamefont {Pitkin}}, \bibinfo {author} {\bibnamefont {paulaltin}}, \bibinfo {author} {\bibfnamefont {J.}~\bibnamefont {Areeda}}, \bibinfo {author} {\bibfnamefont {E.}~\bibnamefont {Quintero}}, \bibinfo {author} {\bibfnamefont {T.~G.}\ \bibnamefont {Badger}}, \bibinfo {author} {\bibfnamefont {L.}~\bibnamefont {Singer}},\ and\ \bibinfo {author} {\bibfnamefont {K.}~\bibnamefont {Leinweber}},\ }\href {https://doi.org/10.5281/zenodo.3598469} {\bibinfo {title} {gwpy/gwpy: 1.0.1}} (\bibinfo {year} {2020})\BibitemShut {NoStop}%
\bibitem [{\citenamefont {Virtanen}\ \emph {et~al.}(2020)\citenamefont {Virtanen}, \citenamefont {Gommers}, \citenamefont {Oliphant}, \citenamefont {Haberland}, \citenamefont {Reddy}, \citenamefont {Cournapeau}, \citenamefont {Burovski}, \citenamefont {Peterson}, \citenamefont {Weckesser}, \citenamefont {Bright}, \citenamefont {{van der Walt}}, \citenamefont {Brett}, \citenamefont {Wilson}, \citenamefont {Millman}, \citenamefont {Mayorov}, \citenamefont {Nelson}, \citenamefont {Jones}, \citenamefont {Kern}, \citenamefont {Larson}, \citenamefont {Carey}, \citenamefont {Polat}, \citenamefont {Feng}, \citenamefont {Moore}, \citenamefont {{VanderPlas}}, \citenamefont {Laxalde}, \citenamefont {Perktold}, \citenamefont {Cimrman}, \citenamefont {Henriksen}, \citenamefont {Quintero}, \citenamefont {Harris}, \citenamefont {Archibald}, \citenamefont {Ribeiro}, \citenamefont {Pedregosa}, \citenamefont {{van Mulbregt}},\ and\ \citenamefont {{SciPy 1.0 Contributors}}}]{2020SciPy-NMeth}%
  \BibitemOpen
  \bibfield  {author} {\bibinfo {author} {\bibfnamefont {P.}~\bibnamefont {Virtanen}}, \bibinfo {author} {\bibfnamefont {R.}~\bibnamefont {Gommers}}, \bibinfo {author} {\bibfnamefont {T.~E.}\ \bibnamefont {Oliphant}}, \bibinfo {author} {\bibfnamefont {M.}~\bibnamefont {Haberland}}, \bibinfo {author} {\bibfnamefont {T.}~\bibnamefont {Reddy}}, \bibinfo {author} {\bibfnamefont {D.}~\bibnamefont {Cournapeau}}, \bibinfo {author} {\bibfnamefont {E.}~\bibnamefont {Burovski}}, \bibinfo {author} {\bibfnamefont {P.}~\bibnamefont {Peterson}}, \bibinfo {author} {\bibfnamefont {W.}~\bibnamefont {Weckesser}}, \bibinfo {author} {\bibfnamefont {J.}~\bibnamefont {Bright}}, \bibinfo {author} {\bibfnamefont {S.~J.}\ \bibnamefont {{van der Walt}}}, \bibinfo {author} {\bibfnamefont {M.}~\bibnamefont {Brett}}, \bibinfo {author} {\bibfnamefont {J.}~\bibnamefont {Wilson}}, \bibinfo {author} {\bibfnamefont {K.~J.}\ \bibnamefont {Millman}}, \bibinfo {author} {\bibfnamefont {N.}~\bibnamefont {Mayorov}}, \bibinfo {author} {\bibfnamefont
  {A.~R.~J.}\ \bibnamefont {Nelson}}, \bibinfo {author} {\bibfnamefont {E.}~\bibnamefont {Jones}}, \bibinfo {author} {\bibfnamefont {R.}~\bibnamefont {Kern}}, \bibinfo {author} {\bibfnamefont {E.}~\bibnamefont {Larson}}, \bibinfo {author} {\bibfnamefont {C.~J.}\ \bibnamefont {Carey}}, \bibinfo {author} {\bibfnamefont {{\.I}.}~\bibnamefont {Polat}}, \bibinfo {author} {\bibfnamefont {Y.}~\bibnamefont {Feng}}, \bibinfo {author} {\bibfnamefont {E.~W.}\ \bibnamefont {Moore}}, \bibinfo {author} {\bibfnamefont {J.}~\bibnamefont {{VanderPlas}}}, \bibinfo {author} {\bibfnamefont {D.}~\bibnamefont {Laxalde}}, \bibinfo {author} {\bibfnamefont {J.}~\bibnamefont {Perktold}}, \bibinfo {author} {\bibfnamefont {R.}~\bibnamefont {Cimrman}}, \bibinfo {author} {\bibfnamefont {I.}~\bibnamefont {Henriksen}}, \bibinfo {author} {\bibfnamefont {E.~A.}\ \bibnamefont {Quintero}}, \bibinfo {author} {\bibfnamefont {C.~R.}\ \bibnamefont {Harris}}, \bibinfo {author} {\bibfnamefont {A.~M.}\ \bibnamefont {Archibald}}, \bibinfo {author}
  {\bibfnamefont {A.~H.}\ \bibnamefont {Ribeiro}}, \bibinfo {author} {\bibfnamefont {F.}~\bibnamefont {Pedregosa}}, \bibinfo {author} {\bibfnamefont {P.}~\bibnamefont {{van Mulbregt}}},\ and\ \bibinfo {author} {\bibnamefont {{SciPy 1.0 Contributors}}},\ }\href {https://doi.org/10.1038/s41592-019-0686-2} {\bibfield  {journal} {\bibinfo  {journal} {Nature Methods}\ }\textbf {\bibinfo {volume} {17}},\ \bibinfo {pages} {261} (\bibinfo {year} {2020})}\BibitemShut {NoStop}%
\bibitem [{\citenamefont {Hunter}(2007)}]{Hunter:2007}%
  \BibitemOpen
  \bibfield  {author} {\bibinfo {author} {\bibfnamefont {J.~D.}\ \bibnamefont {Hunter}},\ }\href {https://doi.org/10.1109/MCSE.2007.55} {\bibfield  {journal} {\bibinfo  {journal} {Computing In Science \& Engineering}\ }\textbf {\bibinfo {volume} {9}},\ \bibinfo {pages} {90} (\bibinfo {year} {2007})}\BibitemShut {NoStop}%
\bibitem [{\citenamefont {pandas~development team}(2020)}]{reback2020pandas}%
  \BibitemOpen
  \bibfield  {author} {\bibinfo {author} {\bibfnamefont {T.}~\bibnamefont {pandas~development team}},\ }\href {https://doi.org/10.5281/zenodo.3509134} {\bibinfo {title} {pandas-dev/pandas: Pandas}} (\bibinfo {year} {2020})\BibitemShut {NoStop}%
\bibitem [{\citenamefont {Harris}\ \emph {et~al.}(2020)\citenamefont {Harris}, \citenamefont {Millman}, \citenamefont {van~der Walt}, \citenamefont {Gommers}, \citenamefont {Virtanen}, \citenamefont {Cournapeau}, \citenamefont {Wieser}, \citenamefont {Taylor}, \citenamefont {Berg}, \citenamefont {Smith}, \citenamefont {Kern}, \citenamefont {Picus}, \citenamefont {Hoyer}, \citenamefont {van Kerkwijk}, \citenamefont {Brett}, \citenamefont {Haldane}, \citenamefont {del R{\'{i}}o}, \citenamefont {Wiebe}, \citenamefont {Peterson}, \citenamefont {G{\'{e}}rard-Marchant}, \citenamefont {Sheppard}, \citenamefont {Reddy}, \citenamefont {Weckesser}, \citenamefont {Abbasi}, \citenamefont {Gohlke},\ and\ \citenamefont {Oliphant}}]{harris2020array}%
  \BibitemOpen
  \bibfield  {author} {\bibinfo {author} {\bibfnamefont {C.~R.}\ \bibnamefont {Harris}}, \bibinfo {author} {\bibfnamefont {K.~J.}\ \bibnamefont {Millman}}, \bibinfo {author} {\bibfnamefont {S.~J.}\ \bibnamefont {van~der Walt}}, \bibinfo {author} {\bibfnamefont {R.}~\bibnamefont {Gommers}}, \bibinfo {author} {\bibfnamefont {P.}~\bibnamefont {Virtanen}}, \bibinfo {author} {\bibfnamefont {D.}~\bibnamefont {Cournapeau}}, \bibinfo {author} {\bibfnamefont {E.}~\bibnamefont {Wieser}}, \bibinfo {author} {\bibfnamefont {J.}~\bibnamefont {Taylor}}, \bibinfo {author} {\bibfnamefont {S.}~\bibnamefont {Berg}}, \bibinfo {author} {\bibfnamefont {N.~J.}\ \bibnamefont {Smith}}, \bibinfo {author} {\bibfnamefont {R.}~\bibnamefont {Kern}}, \bibinfo {author} {\bibfnamefont {M.}~\bibnamefont {Picus}}, \bibinfo {author} {\bibfnamefont {S.}~\bibnamefont {Hoyer}}, \bibinfo {author} {\bibfnamefont {M.~H.}\ \bibnamefont {van Kerkwijk}}, \bibinfo {author} {\bibfnamefont {M.}~\bibnamefont {Brett}}, \bibinfo {author} {\bibfnamefont
  {A.}~\bibnamefont {Haldane}}, \bibinfo {author} {\bibfnamefont {J.~F.}\ \bibnamefont {del R{\'{i}}o}}, \bibinfo {author} {\bibfnamefont {M.}~\bibnamefont {Wiebe}}, \bibinfo {author} {\bibfnamefont {P.}~\bibnamefont {Peterson}}, \bibinfo {author} {\bibfnamefont {P.}~\bibnamefont {G{\'{e}}rard-Marchant}}, \bibinfo {author} {\bibfnamefont {K.}~\bibnamefont {Sheppard}}, \bibinfo {author} {\bibfnamefont {T.}~\bibnamefont {Reddy}}, \bibinfo {author} {\bibfnamefont {W.}~\bibnamefont {Weckesser}}, \bibinfo {author} {\bibfnamefont {H.}~\bibnamefont {Abbasi}}, \bibinfo {author} {\bibfnamefont {C.}~\bibnamefont {Gohlke}},\ and\ \bibinfo {author} {\bibfnamefont {T.~E.}\ \bibnamefont {Oliphant}},\ }\href {https://doi.org/10.1038/s41586-020-2649-2} {\bibfield  {journal} {\bibinfo  {journal} {Nature}\ }\textbf {\bibinfo {volume} {585}},\ \bibinfo {pages} {357} (\bibinfo {year} {2020})}\BibitemShut {NoStop}%
\end{thebibliography}%

\end{document}